\documentclass[12pt,a4paper,aps,preprint,superscriptaddress,nofootinbib]{revtex4-1}
\usepackage[utf8]{inputenc}
\usepackage{graphicx}
\usepackage{amssymb}
\usepackage{textcomp}
\usepackage{amsmath}
\usepackage{tabularx}
\usepackage{bm}
\usepackage{times}
\usepackage{color}
\usepackage{slashed}
\usepackage{multirow}
\usepackage{verbatim}
\usepackage{cancel}
\usepackage{subfigure}
\usepackage{ulem}

\usepackage[colorlinks=true, pdfstartview=FitV, linkcolor=blue, citecolor=blue, urlcolor=blue]{hyperref}
\allowdisplaybreaks[4]

\def\lsim{\mathrel{\raise.3ex\hbox{$<$\kern-.75em\lower1ex\hbox{$\sim$}}}}
\def\gsim{\mathrel{\raise.3ex\hbox{$>$\kern-.75em\lower1ex\hbox{$\sim$}}}}

\def\beq{\begin{equation}}
\def\eeq{\end{equation}}
\def\bea{\begin{eqnarray}}
\def\eea{\end{eqnarray}}

\def \({\left(}
\def \){\right)}
\def \[{\left[}
\def \]{\right]}
\def \l|{\left|}
\def \r|{\right|}

\definecolor{orange}{rgb}{1,0.5,0}

\begin{document}

\title{Reexamining the search for light ALPs at flavor and forward accelerator experiments}

\author{Tong Li}
\email{litong@nankai.edu.cn}
\affiliation{
School of Physics, Nankai University, Tianjin 300071, China
}
\author{Michael A. Schmidt}
\email{m.schmidt@unsw.edu.au}
\affiliation{
Sydney Consortium for Particle Physics and Cosmology,\\
School of Physics, The University of New South Wales, Sydney, New South Wales 2052, Australia
}
\author{Man Yuan}
\email{yuanman@mail.nankai.edu.cn}
\affiliation{
School of Physics, Nankai University, Tianjin 300071, China
}
\preprint{CPPC-2025-05}
\begin{abstract}
The axion-like particle (ALP) is a well-motivated
extension of the Standard Model.
In this work, we revisit the sensitivity of forward accelerator experiments to light long-lived ALPs, and analyze flavor constraints. Our analysis incorporates recent measurements of the rare decays $B\to K + X$ and $K\to \pi +X$, which place stringent bounds on quark flavor violation of a light ALP.
We consider
the complete list of hadronic modes in the calculation of the ALP decay rate and exclusive production channels based on recent improvements.
The analysis includes the discussion of tree-level quark flavor-violating couplings in addition to a universal flavor-conserving ALP coupling to fermions and the electroweak ALP couplings.
Our results demonstrate the complementarity of heavy meson decays and forward accelerator facilities in probing light ALPs. The interplay between two ALP couplings is also investigated.
\end{abstract}

\maketitle

\section{Introduction}

The CP-odd axion-like particles (ALPs) are pseudo-Nambu-Goldstone bosons after the spontaneous symmetry breaking of a global $U(1)$ Peccei-Quinn (PQ) symmetry.
The most well-studied example of an ALP is the QCD axion~\cite{Peccei:1977hh,Peccei:1977ur,Weinberg:1977ma,Wilczek:1977pj} (see the recent review~\cite{DiLuzio:2020wdo}). In general the ALP mass ($m_a$) and the ALP associated symmetry breaking scale (often called decay constant $f_a$) are unrelated~\cite{Dimopoulos:1979pp,Tye:1981zy,Zhitnitsky:1980tq,Dine:1981rt,Holdom:1982ex,Kaplan:1985dv,Srednicki:1985xd,Flynn:1987rs,Kamionkowski:1992mf,Berezhiani:2000gh,Hsu:2004mf,Hook:2014cda,Alonso-Alvarez:2018irt,Hook:2019qoh}. ALP candidates can also appear in many other theoretical models. Their masses span from sub-micro-eV~\cite{Kim:1979if,Shifman:1979if,Dine:1981rt,Zhitnitsky:1980tq,Turner:1989vc} to TeV scale and even beyond~\cite{Rubakov:1997vp,Fukuda:2015ana,Gherghetta:2016fhp,Dimopoulos:2016lvn,Chiang:2016eav,Gaillard:2018xgk,Gherghetta:2020ofz}. Thus, the search for ALPs requires rather different experimental strategies and facilities.

Axion/ALP physics beyond the Standard Model (SM) and the associated detection methods  have attracted significant attention in the particle physics community in recent years.
In this work we revisit flavor constraints and the sensitivity prospects for light long-lived ALPs. For ALPs with $m_a\lesssim \mathcal{O}(1)$ GeV, flavor experiments are sensitive to parameter regions with sizable couplings to SM matter. Recently, the first measurement of rare invisible decay $B^+\to K^+\nu\bar{\nu}$ was performed at Belle II~\cite{Belle-II:2023esi}. The combination
of the inclusive and hadronic tagging analysis results in the branching ratio as
\begin{eqnarray}
{\rm BR}(B^+\to K^+\nu\bar{\nu})_{\rm SD,~ Belle~II}=(2.3\pm 0.7)\times 10^{-5}\;,
\end{eqnarray}
which is 2.7 standard deviations above the SM expectation~\cite{Parrott:2022zte,Becirevic:2023aov}. A long-lived ALP may escape from the detector and mimic SM neutrinos. See Refs.~\cite{Altmannshofer:2023hkn,Altmannshofer:2024kxb,Dai:2024onu,Li:2024thq,Calibbi:2025rpx} for recent ALP interpretations of this Belle II measurement and Refs.~\cite{Berezhiani:1989fs,Berezhiani:1990jj,Berezhiani:1990wn,Ferber:2022rsf} for earlier studies of $B$ meson decays to an ALP. Also, the LHCb, Belle II and BaBar experiments searched for the ALP rare decays to visible final states, e.g. $B\to K^{(\ast)}\mu^+ \mu^-$~\cite{LHCb:2015nkv,LHCb:2016awg,Belle-II:2023ueh} as well as $B^+\to K^+ \gamma\gamma$~\cite{BaBar:2021ich}.
These measurements together with the previous searches~\cite{Belle:2013tnz,Belle:2017oht} provide good opportunities to constrain the flavor-violating interaction between ALP and bottom quark.
Moreover, the NA62 experiment recently provided a measurement of $K^+\to \pi^+ \nu\bar{\nu}$ decay after combining data collected in 2016–2022~\cite{NA62:2024pjp}
\begin{eqnarray}
{\rm BR}(K^+\to \pi^+\nu\bar{\nu})_{\rm NA62}=(13.0^{+3.3}_{-3.0})\times 10^{-11}\;,
\end{eqnarray}
which establishes the detection of this decay mode with a significance above $5\sigma$. Other relevant measurements of visible kaon decay include $K\to \pi \mu^+\mu^-$~\cite{KTEV:2000ngj} and $K\to \pi \gamma\gamma$~\cite{E949:2005qiy,NA62:2023olg,KTeV:2008nqz}. They can also constrain the ALP-strange quark coupling~\cite{Guadagnoli:2025xnt}.

Given their feeble couplings, light ALPs can manifest as long-lived particles (LLPs) and are likely to be detected in accelerator-based forward search experiments (see also the complementary searches for ALP/dark sector particle at colliders, e.g., the LHC and lepton colliders as discussed in Refs.~\cite{Beltran:2023nli,Cheng:2024aco}).
FASER (the ForwArd Search ExpeRiment)~\cite{Feng:2017uoz,FASER:2018ceo,FASER:2018eoc,FASER:2018bac,FASER:2019aik,FASER:2022hcn} is a currently operating experiment designed to explore light particles with extremely weak interactions. The ability to effectively capture forward-traveling particles while minimizing background from collisions makes FASER an ideal environment for studying LLPs. It is located 480 m downstream from the ATLAS interaction point. FASER features a cylindrical decay volume with 1.5 m length and 0.1 m radius. It has been collecting data since 2022 and the total luminosity is expected to reach $\mathcal{L}=300$ fb$^{-1}$ with 13.6 TeV proton-proton collisions at the LHC.
FASER2~\cite{FASER:2018eoc,FASER:2019aik,Feng:2022inv} and FACET (the Forward-Aperture CMS ExTension)~\cite{Boiarska:2019vid,Cerci:2021nlb} are proposed forward search experiments which are expected to be built in the HL-LHC era with total integrated luminosity $\mathcal{L}=3000$ fb$^{-1}$ and center-of-mass (c.m.) energy $\sqrt{s}=14$ TeV. FASER2 as the future updated version of FASER is planned to be built at 620 m downstream of the ATLAS interaction point. For the cylindrical decay volume, its length and radius will be expanded to 5 m and 1 m, respectively. FACET has a decay volume with 18 m length and 0.5 m radius located at 101 m downstream of the CMS interaction point. Compared to FASER and FASER2, it is expected to cover a broader region of LLP parameter space. This is attributed to its longer decay volume and smaller distance to the interaction point, which enables a larger solid angle.

In this work, we aim to investigate the complementarity of heavy meson decays and the forward accelerator facilities for the probe of light ALP.
From a theory perspective, the hadronic decay of ALP remains a long-standing issue. For the regime of $m_a<1$ GeV, the ALP dominantly decays into pions, and the calculation of decay rate can be performed using chiral perturbation theory~\cite{Bauer:2017ris}. In the perturbative region with $m_a>3$ GeV, the dominant ALP decay mode is into heavy quarks. However, a reliable calculation for the intermediate regime has long been absent. Recently, in terms of universal quark couplings of ALP, Refs.~\cite{Aloni:2018vki,Cheng:2021kjg,DallaValleGarcia:2023xhh}~\footnote{Very recently, Ref.~\cite{Ovchynnikov:2025gpx} pointed out an issue with vector mesons and further improved the ALP decay to gluons.}
improved the calculations of the hadronic decay modes by including various hadronic states in ALP decay. This improvement provides us with a good opportunity to re-examine the flavor constraints and long-lived ALP search.
In addition to fermionic ALP couplings, electroweak gauge interactions generate one-loop induced quark flavor-violating ALP coupling and contribute to ALP production processes at flavor facilities and accelerator experiments~\cite{Gavela:2019wzg,FASER:2024bbl,Sun:2025hep}.
Consequently, this motivates a systematic study of correlations among distinct ALP couplings subject to both flavor physics and LLP searches.
This work includes the following characteristics:
\begin{itemize}
\item The complete hadronic decay modes are included in the calculation of ALP decay rate and exclusive heavy meson decays as ALP production processes, by following the procedure in Ref.~\cite{DallaValleGarcia:2023xhh}.
\item The tree-level quark flavor-violating coupling and the coupling of the ALP to the electroweak field strength tensors are taken into account in addition to the universal fermion flavor-conserving coupling.
\item We exhibit the interplay between two ALP couplings in light of flavor constraints and forward accelerator experiments.
\item We demonstrate the complementarity of flavor experiments and the forward accelerator facilities for the probe of light ALP.
\end{itemize}

This paper is organized as follows. In Sec.~\ref{sec:ALP} we introduce the general ALP Lagrangian and discuss the characteristics of its couplings which are essential for the studies of sensitivity reach below. Sec.~\ref{sec:Flav} contains the constraints on the ALP couplings from flavor experiments. In Sec.~\ref{sec:Acc} we
study the sensitivities of CERN accelerator facilities to a light ALP with either fermionic or electroweak gauge coupling. We also investigate the correlation of various ALP couplings and present the discovery sensitivities in the presence of two couplings. Finally, in Sec.~\ref{sec:Con} we summarize our main conclusions.

\section{The general ALP Lagrangian and couplings}
\label{sec:ALP}

We consider a generic massive CP-odd scalar $a$, namely an ALP associated with a global $U(1)$
symmetry spontaneously broken above the electroweak scale. Besides the kinetic term, the most general effective ALP Lagrangian at the UV scale $\Lambda$ is given by~\cite{Georgi:1986df,Bauer:2020jbp,Bauer:2021mvw}
\begin{eqnarray}
\mathcal{L}_{eff}(\Lambda)
&=& C_{GG}{\alpha_s\over 4\pi}{a\over F_a}G^i_{\mu\nu}\tilde{G}_i^{\mu\nu} + C_{WW}{\alpha_2\over 4\pi}{a\over F_a}W^j_{\mu\nu}\tilde{W}_j^{\mu\nu} +
C_{BB}{\alpha_1\over 4\pi}{a\over F_a}B_{\mu\nu}\tilde{B}^{\mu\nu} \nonumber \\
&&+{\partial^\mu a\over F_a}\sum_F \bar{\psi}_F C_F \gamma_\mu \psi_F + C_\Phi {\partial^\mu a\over F_a}(\Phi^\dagger i\overleftrightarrow{D}_\mu \Phi )\;,
\label{eq:effLagrangian}
\end{eqnarray}
where $G_{\mu\nu}^i\ (i=1,\cdots,8)$, $W_{\mu\nu}^j\ (j=1,2,3)$ and $B_{\mu\nu}$ denote the field strength tensors of $SU(3)_c$, $SU(2)_L$ and $U(1)_Y$ gauge fields, respectively, and the dual field strengths are defined as $\tilde{X}^{\mu\nu}\equiv {1\over 2}\epsilon^{\mu\nu\alpha\beta}X_{\alpha\beta}$ with $\epsilon^{0123}=1$. For fermionic couplings, $F=Q,u,d,l,e$ denotes the SM fermion multiplets. Note that the redundant operator ${\partial^\mu a\over F_a}(\Phi^\dagger i\overleftrightarrow{D}_\mu \Phi )$ with $\Phi$ being the SM Higgs doublet here can be removed using the redefinition of fields and a particular choice of operator basis~\cite{Bauer:2021mvw}.

The general ALP-fermion interactions at energy scales $\mu<\Lambda$
can be expressed in terms of the SM fermion mass eigenstates as follows~\cite{Bauer:2021mvw,DallaValleGarcia:2023xhh}
\begin{eqnarray}
\mathcal{L}_{eff}(\mu)
&\supset& {\partial_\mu a\over F_a} [\bar{u}_L k_U (\mu) \gamma^\mu u_L+ \bar{u}_R k_u (\mu) \gamma^\mu u_R + \bar{d}_L k_D (\mu) \gamma^\mu d_L+ \bar{d}_R k_d (\mu) \gamma^\mu d_R \nonumber \\
&&+ \bar{e}_L k_E (\mu)\gamma^\mu e_L+ \bar{e}_R k_e (\mu) \gamma^\mu e_R]\;,
\end{eqnarray}
where $k_U=U_{uL}^\dagger C_Q U_{uL}$, $k_u=U_{uR}^\dagger C_u U_{uR}$, $k_D=U_{dL}^\dagger C_Q U_{dL}$, $k_d=U_{dR}^\dagger C_d U_{dR}$, $k_E=U_{eL}^\dagger C_l U_{eL}$ and $k_e=U_{eR}^\dagger C_e U_{eR}$ and the unitary matrices $U$ relate mass eigenstates to gauge interaction eigenstates.
They diagonalize the
SM Yukawa couplings $U_{fL}^\dagger y_f U_{fR}=y_f^{diag}$ and two of them are related to the CKM matrix $U_{uL}^\dagger U_{dL}=V_{CKM}$.
In the mass basis we then define $k_u=k_d=-k_U=-k_D=K_q$ and $k_e=-k_E=K_l$ for the diagonal components to obtain universal fermion couplings
\begin{eqnarray}
\mathcal{L}_{eff}(\mu)
&\supset&
{\partial_\mu a\over f_a} [c_q(\mu) \sum_q \overline{q} \gamma^\mu \gamma_5 q + c_{\ell}(\mu)\sum_\ell \overline{\ell}\gamma^\mu \gamma_5 \ell]\;.
\label{eq:Lagrangian}
\end{eqnarray}
As the size of $K_i(\Lambda)$ with $i=q, \ell$ is unknown, it is convenient to introduce $c_i(\mu)\equiv K_i(\mu)/K_i(\Lambda)$ with $c_i(\Lambda)=1$ and a new scale $f_a\equiv F_a/K_i(\Lambda)$~\cite{DallaValleGarcia:2023xhh}.
We choose universal couplings $c_q(\Lambda)=c_\ell(\Lambda)=1$ at the scale $\Lambda$.
We further introduce flavor-changing ALP couplings, which
are well-motivated and provide intriguing phenomenological signatures~\cite{MartinCamalich:2020dfe,Bauer:2021mvw,Carmona:2021seb}.
One example are models which explain the hierarchy of charged fermion masses using the Froggatt-Nielsen mechanism~\cite{Froggatt:1978nt} and simultaneously solve the strong CP problem as proposed by Wilczek in 1982~\cite{Wilczek:1982rv} and recently revisited in~\cite{Ema:2016ops,Calibbi:2016hwq,Bjorkeroth:2018ipq,delaVega:2021ugs}. Flavor-violating ALP couplings emerge due to non-universal flavour charges of SM fermions. Generally, if one loosens the assumption of the universality of the PQ current and allows non-universal PQ charges of a flavor symmetry, flavor-violating ALP couplings to SM quarks or leptons arise at tree level~\cite{Wilczek:1982rv,Ema:2016ops,Calibbi:2016hwq,Arias-Aragon:2017eww,Bjorkeroth:2018ipq,delaVega:2021ugs,DiLuzio:2023ndz}.

We thus include tree-level off-diagonal quark couplings in the generic fermion Lagrangian
\begin{eqnarray}
{\partial_\mu a\over f_a} \sum_{q_i\neq q_j} \overline{q}_i \gamma^\mu[\tilde{c}_q+\tilde{c}_q\gamma_5] q_j\;,
\label{eq:Lagrangian}
\end{eqnarray}
where the flavor-violating coefficients are also assumed to be universal for all $q_i\neq q_j$ and
for both vector and axial-vector currents.

After the electroweak symmetry breaking, the interactions between ALP and the physical SM gauge bosons become~\cite{Bauer:2021mvw}
\begin{eqnarray}
\mathcal{L}_{eff}
&\supset& c_{GG}{\alpha_s\over 4\pi}{a\over f_a}G^i_{\mu\nu}\tilde{G}_i^{\mu\nu} + c_{WW}{\alpha\over 2\pi s_\theta^2}{a\over f_a}W_{\mu\nu}\tilde{W}^{\mu\nu} +
c_{\gamma\gamma}{\alpha\over 4\pi}{a\over f_a}F_{\mu\nu}\tilde{F}^{\mu\nu}\nonumber \\
&+&c_{\gamma Z}{\alpha\over 2\pi s_\theta c_\theta}{a\over f_a}F_{\mu\nu}\tilde{Z}^{\mu\nu}+c_{ZZ}{\alpha\over 4\pi s_\theta^2 c_\theta^2} {a\over f_a} Z_{\mu\nu}\tilde{Z}^{\mu\nu}\;,
\end{eqnarray}
where $s_\theta$ ($c_\theta$) is the sine (cosine) of the weak mixing angle $\theta_W$. The coefficients are defined as
\begin{eqnarray}
c_{GG}/f_a&=&C_{GG}/F_a\;,~~~c_{WW}/f_a=C_{WW}/F_a\;,\\
c_{\gamma\gamma}/f_a&=&(C_{WW}+C_{BB})/F_a\;,\\
c_{\gamma Z}/f_a&=&(c_\theta^2 C_{WW}-s_\theta^2 C_{BB})/F_a\;,\\
c_{ZZ}/f_a&=&(c_\theta^4 C_{WW}+s_\theta^4 C_{BB})/F_a\;,
\end{eqnarray}
where the gauge boson coefficients are scale independent due to the choice of normalization. The effective couplings of the ALP to the physical SM particles at both tree level and loop level and the related decay rates can be found in Ref.~\cite{Bauer:2017ris}.

The presence of flavor-conserving couplings and the $c_{WW}$ gauge coupling leads to flavor-violating couplings through the exchange of $W^\pm$ bosons at one-loop level~\cite{Gavela:2019wzg}. The loop induced effective interaction for down-type quarks $d_i\neq d_j\in d,s,b$ is
\begin{eqnarray}
{\partial_\mu a\over f_a} \bar{d}_i G_{d_id_j}\gamma^\mu  P_L d_j\;,
\end{eqnarray}
where the effective coupling at the electroweak scale
$\mu_w=m_t$ is~\cite{Bauer:2020jbp,Bauer:2021mvw,DallaValleGarcia:2023xhh}
\begin{align}
G_{d_id_j}(\mu_w)=-V_{t d_i}^\ast V_{t d_j}\Big\{& -{1\over 6}I_t(\Lambda,\mu_w)+{y_t^2(\mu_w)\over 16\pi^2} \Big[ c_q(\mu_w) \Big(-{\rm ln}~{\mu_w^2 \over m_t^2} + {1\over2} +3 g(x_t)\Big) \; \nonumber \\
&+ {3\alpha\over 2\pi s_\theta^2}\Big( c_{WW}+{3\over 2}(3c_q(\mu_w)+c_\ell(\mu_w))\Big)~g'(x_t)\Big] \Big\}
\label{eq:Gij}
\end{align}
in terms of $x_t=m_t^2/m_W^2$ and
\begin{eqnarray}
&&I_t(\Lambda,\mu_w)=\int_{\mu_w}^\Lambda {d\mu\over \mu} {3y_t^2(\mu)\over 88\pi^2} c_q(\mu)\;,\\
&&g(x)={1-x+ \ln x\over (1-x)^2}\;,~~g'(x)={1-x+x\,\ln x\over (1-x)^2}\;.
\end{eqnarray}
After combining with the tree-level off-diagonal couplings, the complete flavor-violating couplings of down-type quarks are given by
\begin{eqnarray}
\partial_\mu a \bar{d}_i \gamma^\mu \Big[\Big({\tilde{c}_q\over f_a}+{G_{d_i d_j}\over 2f_a}\Big) + \Big({\tilde{c}_q\over f_a}-{G_{d_i d_j}\over 2f_a}\Big)\gamma_5 \Big]d_j\;.
\end{eqnarray}
They depend on three key couplings $c_q=c_\ell$, $c_{WW}$ and $\tilde{c}_q$. These interactions in particular result in flavor-violating ALP decays into lighter quarks $a\to d_i \bar{d}_j+\bar{d}_i d_j$, with the following decay rate for $m_a>m_{d_i}+m_{d_j}$
\begin{eqnarray}
\Gamma_{a\to d_id_j}&=&2\Gamma(a\to d_i \bar{d}_j) \\
&=&{N_c\over 4\pi m_a^3 } \Big\{\Big|{\tilde{c}_q\over f_a}+{G_{d_i d_j}\over 2f_a}\Big|^2\Big[m_a^2(m_{d_i}-m_{d_j})^2-(m_{d_i}^2-m_{d_j}^2)^2\Big]\nonumber \\
&&+\Big|{\tilde{c}_q\over f_a}-{G_{d_i d_j}\over 2f_a}\Big|^2\Big[m_a^2(m_{d_i}+m_{d_j})^2-(m_{d_i}^2-m_{d_j}^2)^2\Big] \Big\}
\times
\lambda^{1/2}(m_a^2,m_{d_i}^2,m_{d_j}^2)\\
&=& {N_c m_a m_{d_i}^2\over 4\pi } \Big(\Big|{\tilde{c}_q\over f_a}+{G_{d_i d_j}\over 2f_a}\Big|^2+\Big|{\tilde{c}_q\over f_a}-{G_{d_i d_j}\over 2f_a}\Big|^2\Big) \Big(1-{m_{d_i}^2\over m_a^2}\Big)^2 ~~{\rm when}~m_{d_j}=0\;,
\label{eq:2body-decay}
\end{eqnarray}
where $\lambda(x,y,z)=x^2+y^2+z^2-2xy-2xz-2yz$ denotes the K\"all\'en function.
For up-type quarks, we only include the tree-level flavor-violating coupling and neglect the loop-level one ($G_{u_i u_j}\approx 0$) which is suppressed by a down-type quark Yukawa coupling.

The renormalization group (RG) equations in Ref.~\cite{Bauer:2020jbp} can be solved for the flavor-conserving fermion couplings from $\Lambda$ to the electroweak scale $\mu_w$ and from $\mu_w$ to the ALP mass scale. The RG evolutions are implemented in the mathematica-based code SensCalc~\cite{DallaValleGarcia:2023xhh,SensCalc}. We adopt and modify SensCalc
to include the complete couplings described above in the calculation of ALP decay rate and lifetime.
The gauge coupling $c_{WW}$ is invariant under RG evolution, We further neglect the RG evolution of the tree-level flavor-violating coupling $c_{\tilde{q}}$, because their RG evolution is suppressed by small SM Yukawa couplings or electroweak interactions.

\begin{figure}[htb!]
\centering
\includegraphics[width=0.485\textwidth]{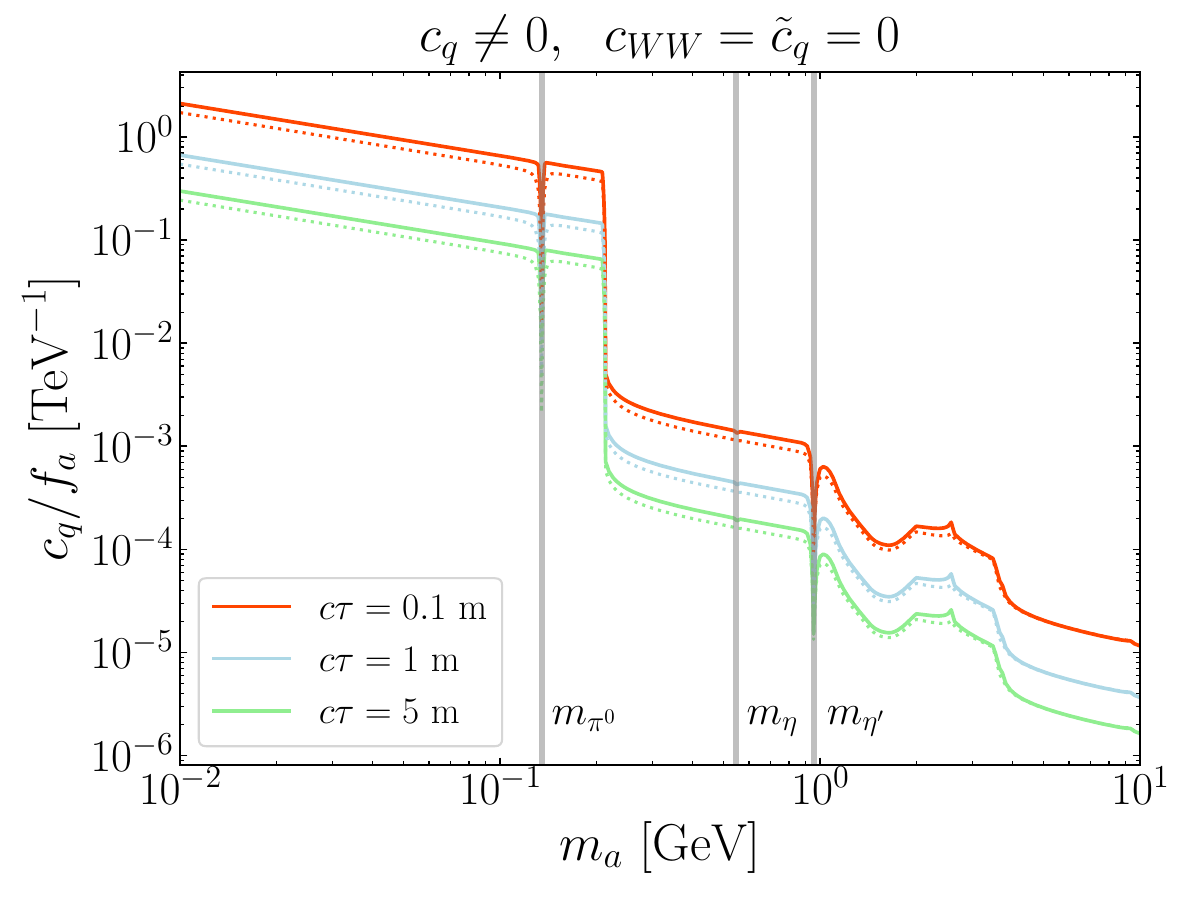}
\includegraphics[width=0.485\textwidth]{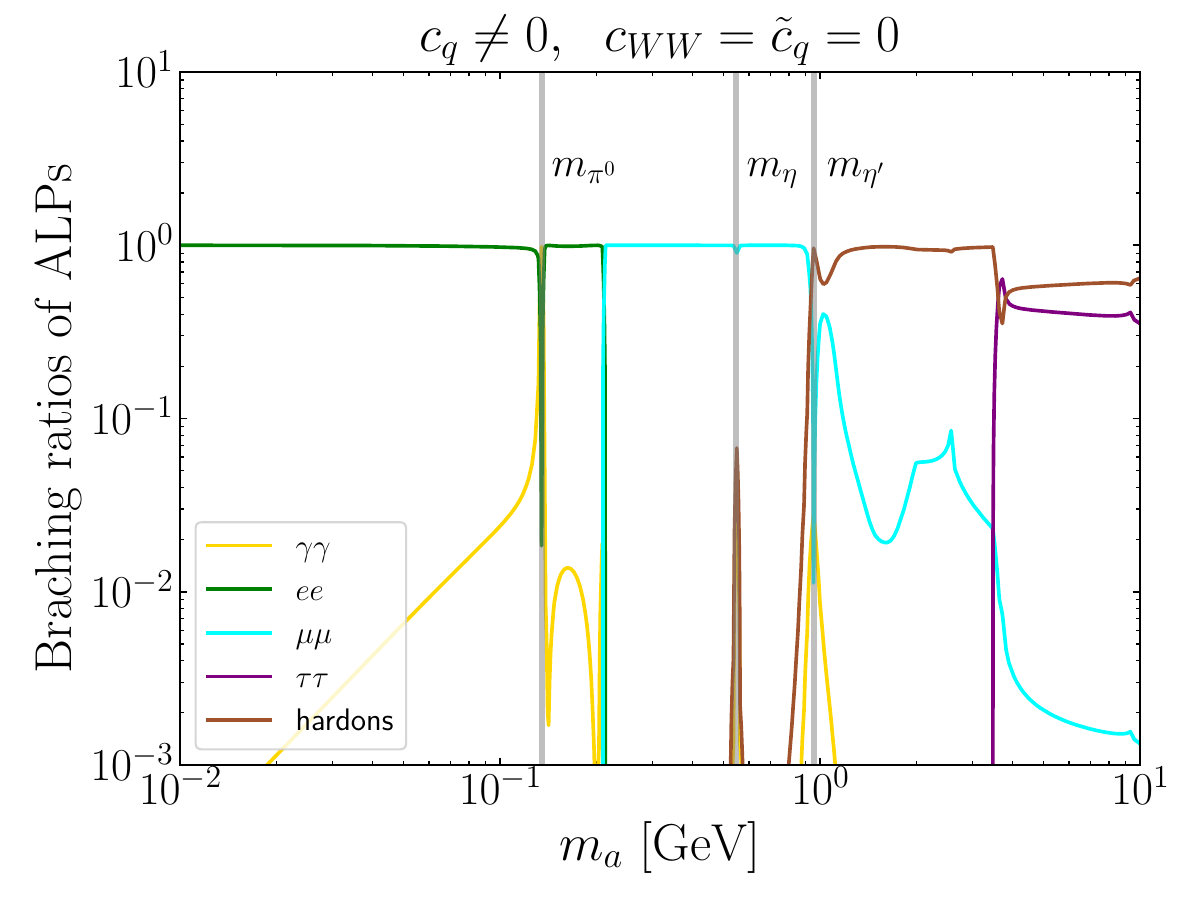}\\
\includegraphics[width=0.485\textwidth]{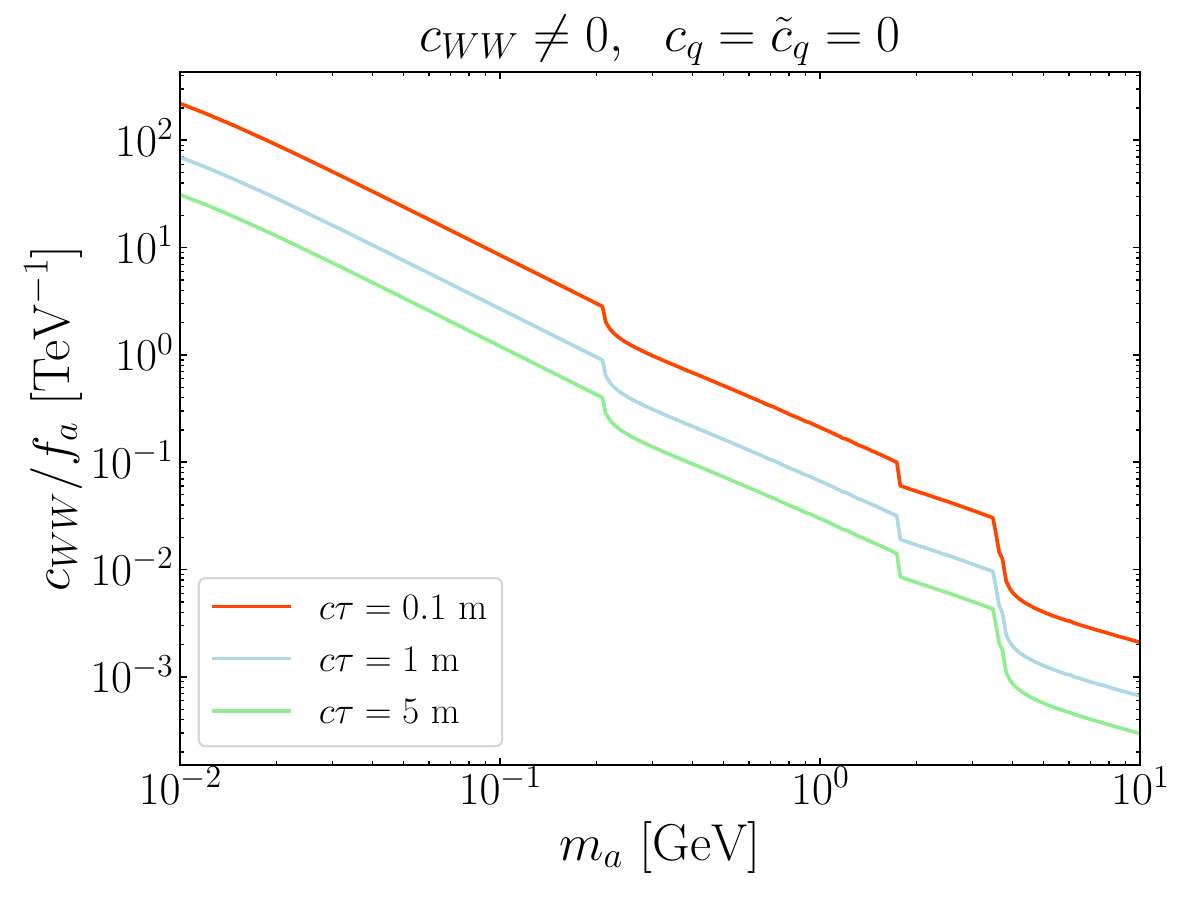}
\includegraphics[width=0.485\textwidth]{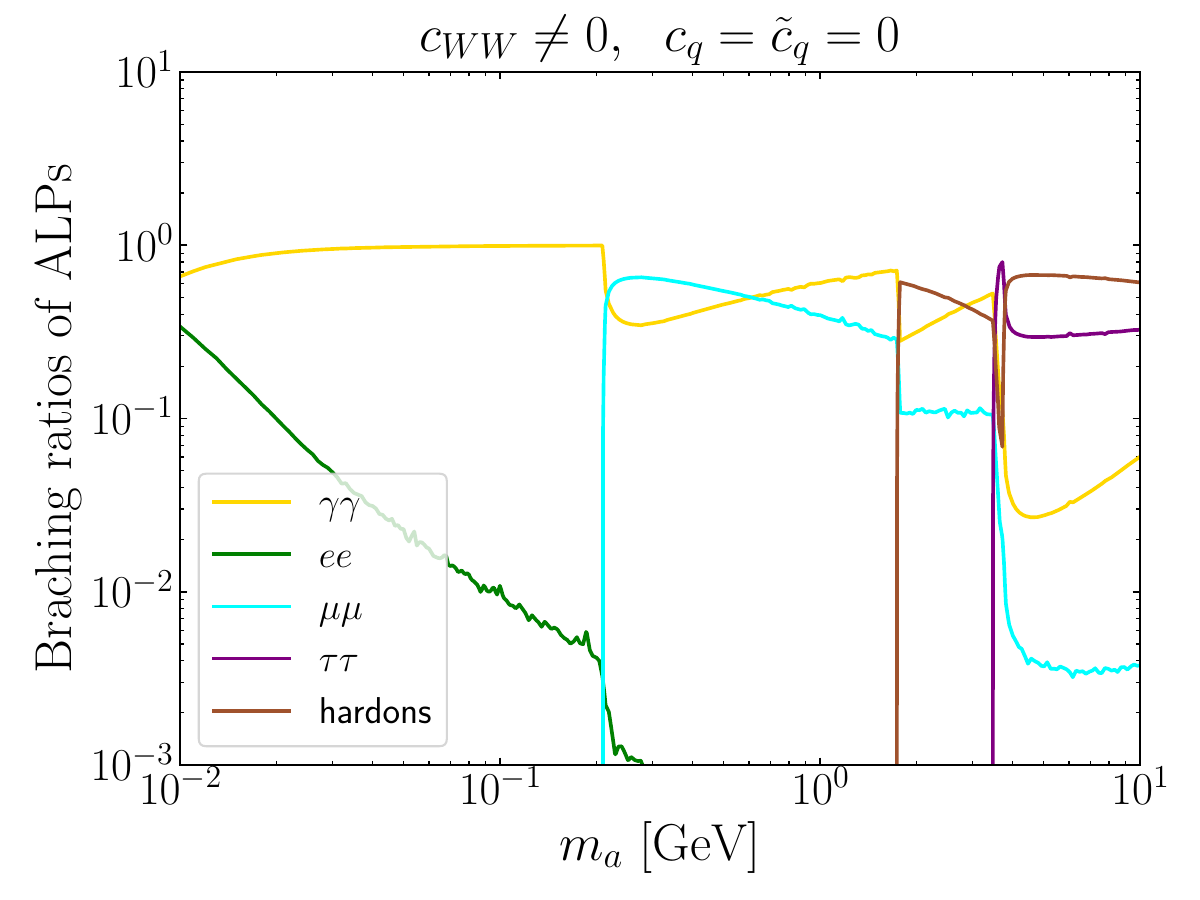}\\
\includegraphics[width=0.485\textwidth]{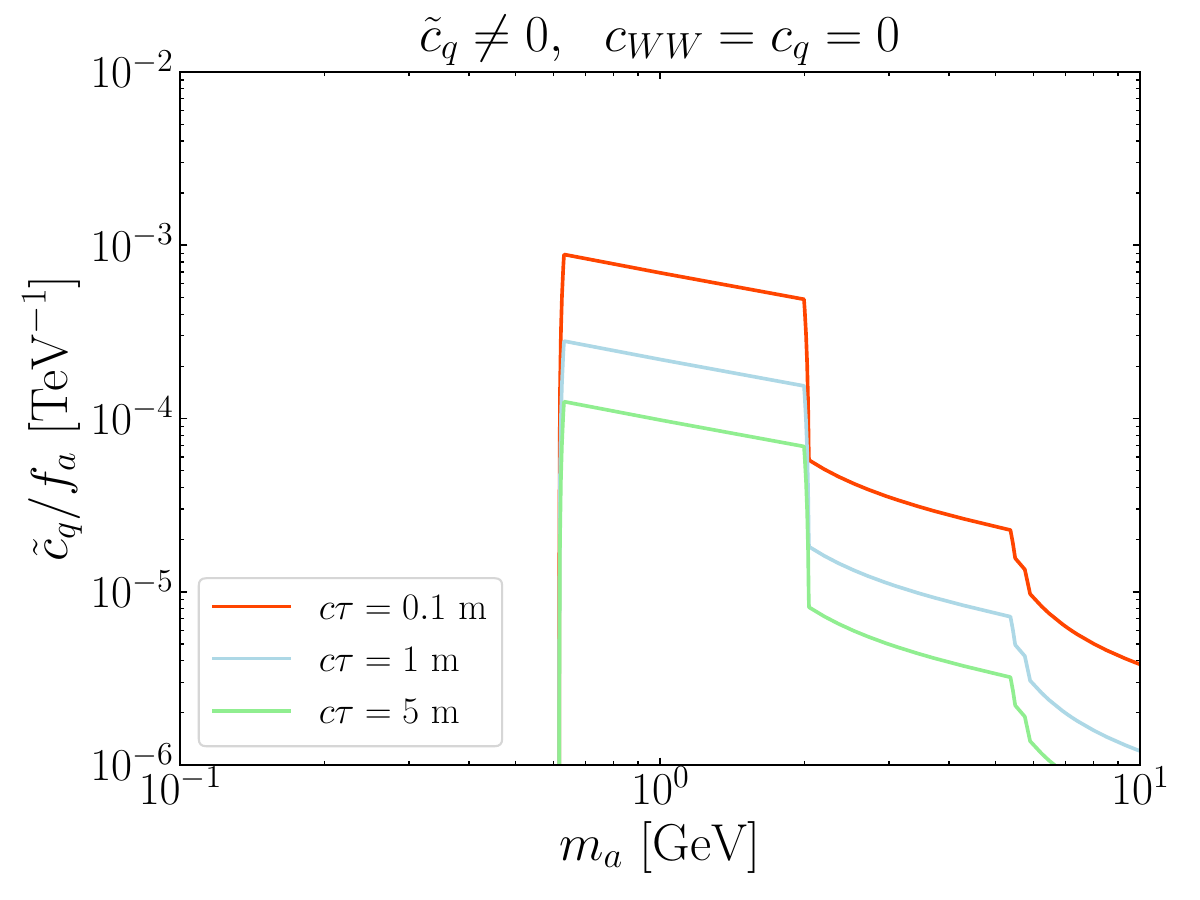}
\includegraphics[width=0.485\textwidth]{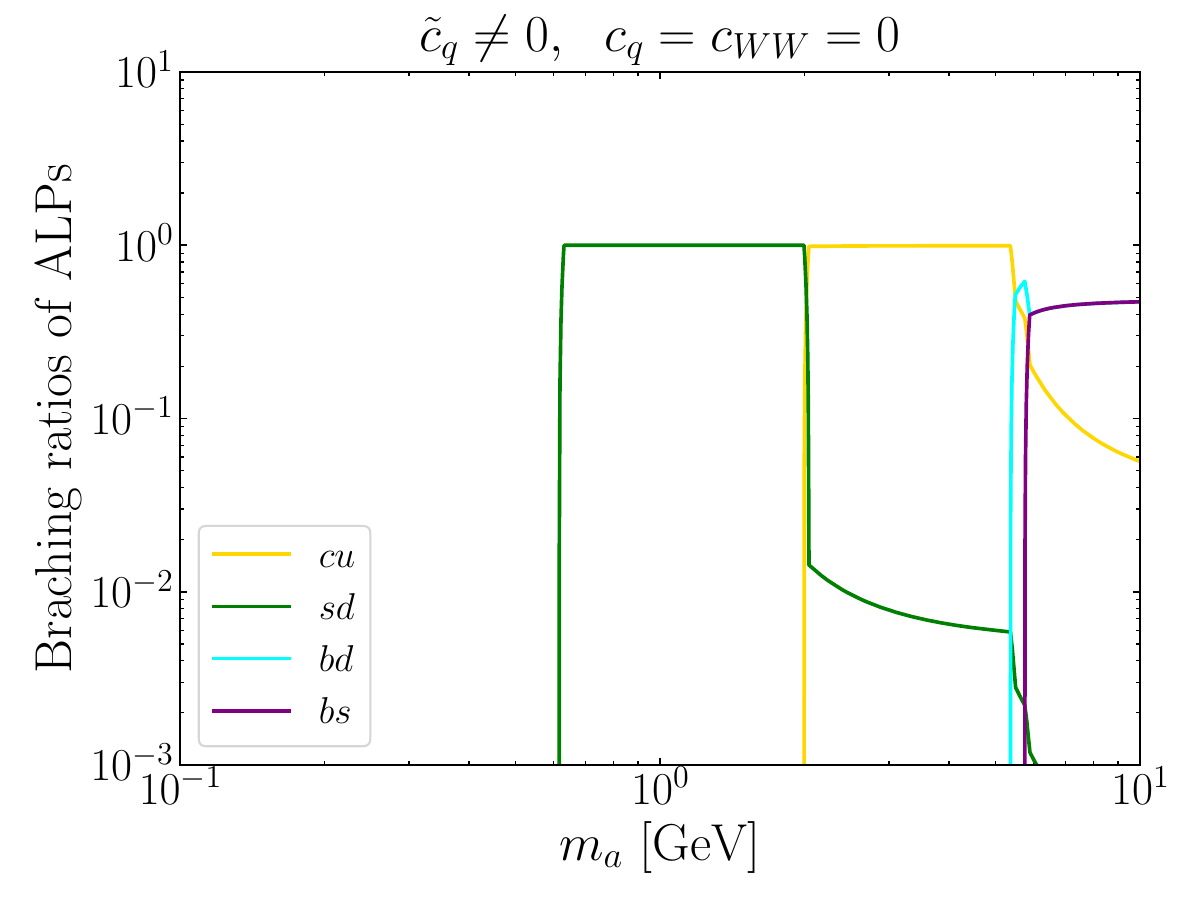}
\caption{Left panels: the ALP couplings versus $m_a$ for decay lengths $c\tau=0.1~{\rm m}$ (red), $1~{\rm m}$ (blue), $5~{\rm m}$ (green). Right panels: ALP decay branching ratios as a function of $m_a$. We consider three choices for ALP couplings: $c_q\neq0,~c_{WW}=\tilde{c}_q=0$ (top), $c_{WW}\neq0,~c_q=\tilde{c}_q=0$ (middle) and $\tilde{c}_q\neq0,~c_q=c_{WW}=0$ (bottom).
The ALP couplings $\tilde{c}_q$ and $c_{WW}$ are RG invariant. For $c_q\neq 0$ we show the results of $\Lambda=10^9$ GeV.
}
\label{fig:BR}
\end{figure}

In Fig.~\ref{fig:BR} we show the contours of ALP couplings versus $m_a$ for lifetimes $c\tau=0.1~{\rm m}$, $1~{\rm m}$, $5~{\rm m}$ (left panels) and the ALP decay branching ratios (right panels). Three choices of ALP couplings are considered: $c_q\neq0,~c_{WW}=\tilde{c}_q=0$ (top panels), $c_{WW}\neq0,~c_q=\tilde{c}_q=0$ (middle panels), and $\tilde{c}_q\neq0,~c_q=c_{WW}=0$ (bottom panels). For the $c_q\neq 0$ case we only show the results of $\Lambda=10^9~{\rm GeV}$ for illustration. The three settings from top to bottom in Fig.~\ref{fig:BR} correspond to the existence of both tree-level and loop-level couplings, only loop-level couplings and only tree-level couplings, respectively.
As one can see, the ALP couplings exhibit significant differences across these three cases for a fixed lifetime and decay product characteristics.
We will analyze the flavor constraints and the sensitivity of forward facilities to ALP couplings for these three choices below.

\section{Flavor constraints on ALP couplings}
\label{sec:Flav}

In this section we investigate the flavor constraints on ALP couplings in light of heavy meson decays and the mixing of neutral mesons.

\subsection{Heavy meson decays}

We first consider the emittance of an ALP in meson decay $M\to M'~a$. The branching fractions of meson decay to invisible and visible final states are given by~\cite{Greljo:2024evt,Calibbi:2024rcm}
\begin{eqnarray}
{\rm BR}(M\to M' + {\rm inv.}) &=& {\rm BR}(M\to M' + a) {\rm BR}(a\to \sum {\rm vis.}){\rm exp}(-L/L_a)\;,
\label{eq:Minv}\\
{\rm BR}(M\to M' + {\rm vis.}) &=& {\rm BR}(M\to M' + a) {\rm BR}(a\to {\rm vis.})\Big[1-{\rm exp}(-L/L_a)\Big]\;,
\label{eq:Mvis}
\end{eqnarray}
where $L\sim 1$ m denotes the typical radius of the detector and $L_a$ is the mean flight length of the boosted ALP with
\begin{eqnarray}
L_a={p_a c\over m_a \Gamma_a}\;,~~p_a={\lambda^{1/2}(m_{M}^2,m_{M'}^2,m_a^2)\over 2m_{M}}\;.
\end{eqnarray}
The rates of on-shell ALP production from heavy meson decay are
\begin{eqnarray}
\Gamma(M\to Pa)&=&{m_{M}^3\over 16\pi}\Big|{\tilde{c}_q\over f_a}+{G_{d_i d_j}\over 2f_a}\Big|^2 \Big(1-\frac{m_P^2}{m_M^2}\Big)^2 f_0^2(m_a^2)\lambda^{1/2}\left(1,{m_P^2\over m_M^2},{m_a^2\over m_M^2}\right)\;,
\label{eq:Pmesonrate}\\
\Gamma(M \to V a) &=&
\frac{m_M^3}{16\pi}\Big|{\tilde{c}_q\over f_a}-{G_{d_i d_j}\over 2f_a}\Big|^2A_0^2(m_a^2)\lambda^{3/2}\Big(1,{m_V^2\over m_M^2},{m_a^2\over m_M^2}\Big)\;,
\label{eq:Vmesonrate}
\end{eqnarray}
where $M'=P, V$ denotes a pseudo-scalar and vector meson, respectively. For decay products of $\pi^0, \rho^0\sim (u\bar{u}-d\bar{d})/\sqrt{2}$, there is an additional overall factor of $1/2$ in the above decay rates.
For $K\to \pi a$ decay, in addition to the contribution from flavor-violating ALP-fermion couplings discussed before, we include the complete transition amplitude from the chiral representation of the effective weak Lagrangian~\cite{Bauer:2021wjo,Bauer:2021mvw}.

\begin{table}[tbp!]
\centering
\begin{tabular}{c|c}
\hline
\hline
    $B$ invisible decays & measurements \\
    \hline
    \hline
    ${\rm BR}(B^+\to K^+ + {\rm inv.})_{\rm NP}$ & $=(1.9\pm 0.7)\times 10^{-5}$~\cite{He:2022ljo,Belle-II:2023esi,Becirevic:2023aov} or Fig.~3 in Ref.~\cite{Fridell:2023ssf}
    \\
    ${\rm BR}(B^+\to K^{\ast +} + {\rm inv.})_{\rm NP}$ & $<3.1\times 10^{-5}$~\cite{Belle:2013tnz,He:2022ljo}  \\
    ${\rm BR}(B^0\to K^{0} + {\rm inv.})_{\rm NP}$ & $<2.3\times 10^{-5}$~\cite{Belle:2017oht,He:2022ljo} \\
    ${\rm BR}(B^0\to K^{\ast 0} + {\rm inv.})_{\rm NP}$ & $<1.0\times 10^{-5}$~\cite{Belle:2017oht,He:2022ljo} \\
    \hline
    ${\rm BR}(B^+\to \pi^{+} + {\rm inv.})_{\rm NP}$ & $<1.4\times 10^{-5}$~\cite{Belle:2017oht,He:2022ljo} \\
    ${\rm BR}(B^+\to \rho^{+} + {\rm inv.})_{\rm NP}$ & $<3.0\times 10^{-5}$~\cite{Belle:2017oht,He:2022ljo} \\
    ${\rm BR}(B^0\to \pi^{0} + {\rm inv.})_{\rm NP}$ & $<8.9\times 10^{-6}$~\cite{Belle:2017oht,He:2022ljo} \\
    ${\rm BR}(B^0\to \rho^{0} + {\rm inv.})_{\rm NP}$ & $<4.0\times 10^{-5}$~\cite{Belle:2017oht,He:2022ljo} \\
    \hline
    \hline
\end{tabular}
\caption{The invisible decay modes of $B$ mesons and the corresponding measurement or bounds on NP contribution.
The upper bounds have been obtained by subtracting the lower bound of the SM prediction from the experimental branching ratio following the procedure in Refs.~\cite{He:2022ljo,He:2023bnk}.
}
\label{tab:semi-invisible_BR}
\end{table}

For $B$ invisible decays, after subtracting the SM contributions, Ref.~\cite{He:2022ljo} derived upper limits on non-interfering new physics (NP) contributions which are reproduced in Table~\ref{tab:semi-invisible_BR} together with the measurement of ${\rm BR}(B^+\to K^+ + {\rm inv.})$~\cite{Belle-II:2023esi}. This analysis only considered the branching ratio and did not take into account the missing invariant mass distribution of the neutrino pair. More recently, Ref.~\cite{Fridell:2023ssf} performed a likelihood analysis taking into account the missing invariant mass distribution extracted from the post-fit distribution of the recent $B^+\to K^+ + \mathrm{inv}.$ measurement~\cite{Belle-II:2023esi} and the BaBar search for $B\to K+\mathrm{inv}.$~\cite{BaBar:2013npw}. They considered a two-body decay $B^+\to K^+X$ with an invisible particle $X$. Their figure 3 shows that the excess can be explained for $m_X\approx 2$ GeV and branching fraction $(0.5-0.9)\times 10^{-5}$ at $1\sigma$. We will present the constraints on a long-lived ALP from both the non-fitted and the fitted results.

LHCb and Belle II searched for displaced vertices in the decay $B\to K^{(*)} a (\to \mu^+\mu^-)$~\cite{LHCb:2015nkv,LHCb:2016awg,Belle-II:2023ueh}. The Belle II analysis also places constraints on $B\to K^{(*)}e^+e^-$. See also the recent phenomenological study of the Belle II sensitivity~\cite{Ferber:2022rsf}. BaBar searched for ALPs decaying to a pair of photons and the constraints are shown in figures 3 and 4 of Ref.~\cite{BaBar:2021ich}. The most stringent constraint is placed by the LHCb search for $B^+\to K^+a(\to\mu^+\mu^-)$ with a displaced vertex~\cite{LHCb:2016awg} which presents constraints on the branching ratio $\mathrm{BR}(B^+\to K^+ a) \mathrm{BR}(a\to \mu^+\mu^-)$ as a function of the mass $m_a$ and lifetime $\tau_a$.
By including the decay lengths of ALP in Eq.~(\ref{eq:Mvis}), we recast the experimental limit of $B\to K^{(\ast)}a(\to \mu^+\mu^-)$ on a bound on the ALP parameters.
We also derive the bounds on NP contribution for $K\to \pi$ rare decays in Table~\ref{tab:semi-invisible_BRK}. The upper bounds have been obtained by subtracting
the lower limit of the 90\% C.L. SM range from the 90\% experimental upper bound. We ignore the observables with large error of SM prediction.

\begin{table}[tb!]
\centering
\begin{tabular}{c|c|c}
\hline
\hline
    $K$ decays & SM predictions & NP measurements \\
    \hline
    \hline
    ${\rm BR}(K^+\to \pi^+ + \nu\bar{\nu})_{\rm exp}=(13.0^{+3.3}_{-3.0})\times 10^{-11}$~\cite{NA62:2024pjp} & $=(7.86\pm 0.61)\times 10^{-11}$~\cite{DAmbrosio:2022kvb} & $=(5.14\pm 2.54)\times 10^{-11}$ \\
    ${\rm BR}(K_L\to \pi^0 + \nu\bar{\nu})_{\rm exp}<3.0\times 10^{-9}$~\cite{KOTO:2018dsc} & $=(3.41\pm 0.45)\times 10^{-11}$~\cite{Hou:2024vyw} & $<2.97\times 10^{-9}$ \\
        \hline
    ${\rm BR}(K_L\to \pi^0 + \mu^+\mu^-)_{\rm exp}<3.8\times 10^{-10}$~\cite{KTEV:2000ngj} & $=(1.39^{+0.27}_{-0.25})\times 10^{-11}$~\cite{Buras:2022qip} & $<3.7\times 10^{-10}$ \\
        \hline
    ${\rm BR}(K^+\to \pi^+\gamma\gamma)_{\rm exp}<8.3\times 10^{-9}$~\cite{E949:2005qiy} & $=6.1\times 10^{-9}$~\cite{Gerard:2005yk} & $<2.2\times 10^{-9}$ \\
    for $m_{\gamma\gamma}<108~{\rm MeV}$ & & \\
    \hline
   ${\rm BR}(K_L\to \pi^0\gamma\gamma)_{\rm exp}=(1.29\pm 0.058)\times 10^{-6}$~\cite{KTeV:2008nqz}  & $=1.12\times 10^{-6}$~\cite{DAmbrosio:1996kjn} & $=(0.17\pm 0.058)\times 10^{-6}$\\
   for $160<m_{\gamma\gamma}<363~{\rm MeV}$ & & \\
    \hline
    \hline
\end{tabular}
\caption{The decay modes of $K$ mesons, SM predictions and the corresponding measurement or bounds on NP contribution.
}
\label{tab:semi-invisible_BRK}
\end{table}

\subsection{Neutral meson mixing}

Finally, we consider constraints from $B$ and $K$ meson mixing\footnote{We do not consider $D$ meson mixing because of its large theoretical uncertainty and the comparatively large error of the measured mass difference $\Delta m_D = (9.97\pm1.16) \,\mathrm{ps}^{-1}$.} following the discussion in Refs.~\cite{MartinCamalich:2020dfe,Bauer:2021mvw,Li:2024thq}.
The kaon mass difference in terms of the couplings $\tilde c_q$ and $G_{sd}$ is given by $\Delta m_K = 2\, \mathrm{Re}(M_{12})$~\cite{MartinCamalich:2020dfe}
\begin{equation}
\begin{aligned}
    \frac{\Delta m_K}{m_K} & =
    \mathrm{Re}[(2\tilde c_q + G_{sd})^2]  \frac{f_K^2}{4f_a^2} \left(1-\frac{m_\pi^2}{m_K^2}\right)^2    \left[ \frac{m_K^2}{32\pi^2 f_K^2}\left(I^\prime_0\left(\frac{m_\pi^2}{m_K^2}\right) + \frac13 I^\prime_0\left(\frac{m_\eta^2}{m_K^2}\right)\right)    \right]
   \\
    & +
    \mathrm{Re}[(2\tilde c_q - G_{sd})^2]
    \frac{f_K^2}{4f_a^2}  \left[1+\frac83 \frac{m_K^2}{16\pi^2 f_K^2}\right] \;.
\end{aligned}
\end{equation}
with $I^\prime_0(z) = 2 -z \ln z - (1-z) \ln(|z-1|)$. This has to be compared with the SM prediction $\Delta m_K^{\rm SM}=5.8(0.6)_{\rm stat}(2.3)_{\rm syst}\times 10^{-12}$ MeV~\cite{Wang:2022lfq}  and the experimental measurement $\Delta m_K^{\rm exp} = (3.484\pm0.006)\times 10^{-12}$ MeV~\cite{ParticleDataGroup:2024cfk}.

\begin{table}\centering
    \begin{tabular}{l|c|c}
    \hline\hline
    & SM prediction & experiment  \\
    \hline\hline
    $\Delta M_{B_d} [\mathrm{ps}^{-1}]$ & $0.533^{+0.022}_{-0.036}$ & $0.5064\pm0.0019$
    \\\hline
    $\Delta M_{B_s} [\mathrm{ps}^{-1}]$ &  $18.4^{+0.7}_{-1.2}$ & $17.7656\pm0.0057$ \\
    \hline\hline
    \end{tabular}

    \vspace{5mm}

    \begin{tabular}{c|c|c|c|c}
    \hline\hline
    $i$ &  2& 3 & 4  &5 \\\hline
    \hline
        $f_{B_d}^2B_{B_d}^{(i)} [\mathrm{GeV}^2]$  &
        $0.0288\pm0.0013$ & $0.0281\pm0.0020$ & $0.0387\pm0.0015$  & $0.0361\pm0.0014$ \\\hline
        $f_{B_s}^2B_{B_s}^{(i)} [\mathrm{GeV}^2]$ &
        $0.0441\pm0.0017$ & $0.0454\pm0.0027$ & $0.0544\pm0.0019$ & $0.0507\pm0.0017$ \\
    \hline\hline
    \end{tabular}
\caption{Top: SM predictions~\cite{DiLuzio:2019jyq} and experimental measurements~\cite{HFLAV:2016hnz,LHCb:2021moh} for $B_q$ mass differences.
Bottom: Hadronic parameters $f_{B_q}^2B_{B_q}^{(i)}$ for the relevant operators~\cite{DiLuzio:2019jyq}.
}
\label{tab:Mixing}
\end{table}

The $B_q$ meson mass difference is given by~\cite{Bauer:2021mvw,Li:2024thq}
\begin{equation}
\begin{aligned}
\Delta M_q=
\Bigg| -\frac{(\lambda_t^q)^2}{|\lambda_t^q|^2} \Delta M_q^{\rm SM}
&+ f_{B_q}^2 m_{B_q}\Bigg[  \left(C^q_2(\mu_b)+\tilde C^q_2(\mu_b) \right)\eta^q_2(\mu_b) B_{B_q}^{(2)}(\mu_b)
\\&
+  \left(C^q_3(\mu_b)+\tilde C^q_3(\mu_b) \right)\eta^q_3(\mu_b) B_{B_q}^{(3)}(\mu_b)
\\ &
+ C^q_4(\mu_b) \eta^q_4(\mu_b) B_{B_q}^{(4)}(\mu_b)
+ C^q_5(\mu_b) \eta^q_5(\mu_b) B_{B_q}^{(5)}(\mu_b)
\Bigg]\Bigg|
\;,
\label{eq:DelMq}
\end{aligned}
\end{equation}
where the first term denotes the SM contribution with $\lambda_t^q\equiv V_{tq}^* V_{tb}$, $C^q_i$ and $\tilde C^q_i$ describe the ALP contributions to the different Wilson coefficients which are given in Ref.~\cite{Bauer:2021mvw}, $f_{B_q}$ is the decay constant of $B_q$, $m_{B_q}$ the $B_q$ meson mass, $B_{B_q}^{(i)}$ are hadronic parameters taken from Ref.~\cite{DiLuzio:2019jyq}
which are reproduced in Table~\ref{tab:Mixing} (bottom), and the $\eta_i^q(\mu_b)$ normalization factors are defined in Ref.~\cite{Bagger:1997gg,Bauer:2021mvw}. The SM prediction and the experimental measurements are collected in Table~\ref{tab:Mixing} (top).

For both meson mass differences the theoretical error of the SM predictions is much larger than the experimental error. In the following numerical analysis, we add the theoretical error to the experimental error in quadrature to take the large theoretical error into account.

\section{ALP sensitivity reach at forward facilities}
\label{sec:Acc}

A very weak ALP coupling can result in a long-lived ALP which can be searched for in forward accelerator facilities such as FASER, FASER2 and FACET. Throughout this section, we revisit the sensitivity of forward experiments to both the ALP quark coupling and ALP coupling to the electroweak field strength tensors.

The major production mechanisms of ALP at FASER include the decays of mesons as well as the mixing between ALP and pseudo-scalar mesons. We use the SensCalc code~\cite{SensCalc} to calculate the ALP lifetime and decay branching fractions. These results are then passed to the
FORESEE package~\cite{Kling:2021fwx} in which the meson production probabilities are computed. In the FORESEE package, we modify the ALP production rates by including various exclusive decay modes of $B$, $K$ and $D$ mesons
and the RG evolutions of ALP fermion couplings down to 2 GeV scale. Note that the $B$ meson can also decay into scalar or pseudo-vector meson in addition to the pseudo-scalar and vector mesons mentioned in Sec.~\ref{sec:Flav}. Ref.~\cite{DallaValleGarcia:2023xhh} pointed out that including the additional decay modes would increase the total production probability four times. Thus, besides pseudo-scalar mesons $P=\pi,~K$ and vector mesons $V=\rho, K^\ast (892), K^\ast (1410), K^\ast (1680)$, we also include the $B$ meson decays to scalar states $S=K_0^\ast (700)$, $K_0^\ast (1430)$ and pseudo-vectors $P_v=K_1(1270)$, $K_1(1400)$ with the partial decay widths as follows
\begin{eqnarray}
\Gamma(B\to S~ a) &=& {(m_B^2-m_S^2-m_a^2)^2\over 16\pi~m_B} \Big|{\tilde{c}_q\over f_a}-{G_{bs}\over 2f_a}\Big|^2 f_+^{2}(m_a^2)\lambda^{1/2}\left(1,{m_{S}^2\over m_B^2},{m_a^2\over m_B^2}\right)\;, \\
\Gamma(B\to P_v~ a) &=& {m_B^3\over 16\pi} \Big|{\tilde{c}_q\over f_a}+{G_{bs}\over 2f_a}\Big|^2 V_0^{2}(m_a^2)~\lambda^{3/2}\left(1,{m_{P_v}^2\over m_B^2},{m_a^2\over m_B^2}\right)\;,
\end{eqnarray}
where the form factors $f_+$ and $V_0$ together with $f_0$ and $A_0$ in Eqs.~(\ref{eq:Pmesonrate}, \ref{eq:Vmesonrate}) can be found in Refs.~\cite{Ball:2004ye,Sun:2010nv,Ball:2004rg,Lu:2011jm,Hatanaka:2009gb,Bashiry:2009wq,Boiarska:2019jym}.
The mixing with light pseudo-scalar mesons ($\pi^0$, $\eta$ and $\eta'$) are also implemented by following the approach of matching with the chiral perturbation theory Lagrangian described in Ref.~\cite{DallaValleGarcia:2023xhh}. We specify the allowed ALP decay products to be $\gamma\gamma$, $e^+e^-$, $\mu^+\mu^-$ and hadrons. Then, FORESEE provides the number of signal events which pass the selection
criteria in forward detectors. Assuming zero observed events and mean background, the sensitivity is estimated by requiring 2.3 signal events to
accommodate 90\% C.L. interval for the Poisson signal mean.

\begin{figure}[tbp!]
\begin{center}
\includegraphics[width=0.8\textwidth]{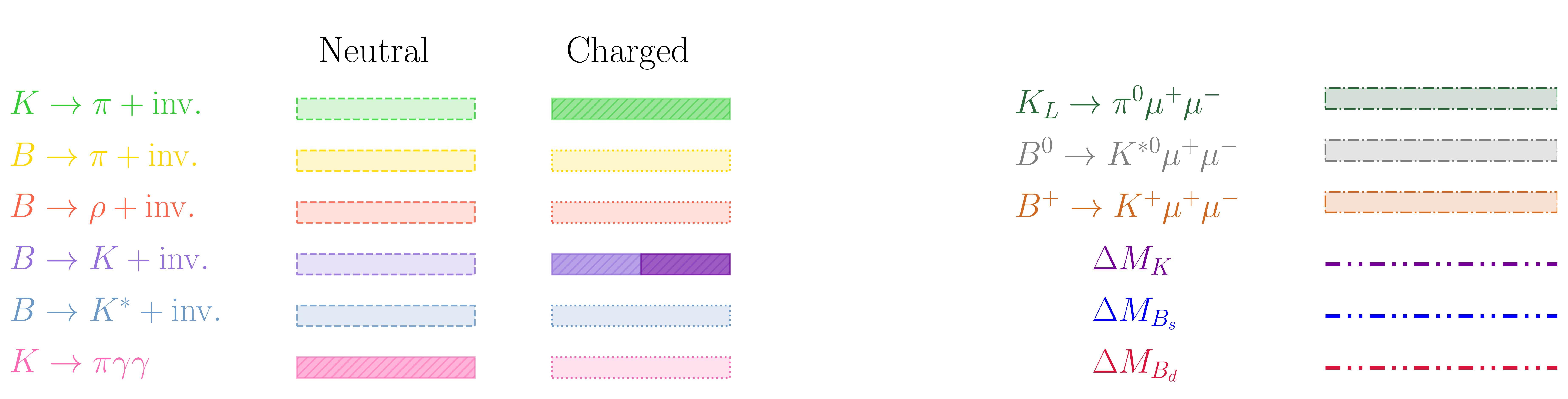}\\
\includegraphics[width=0.475\textwidth]{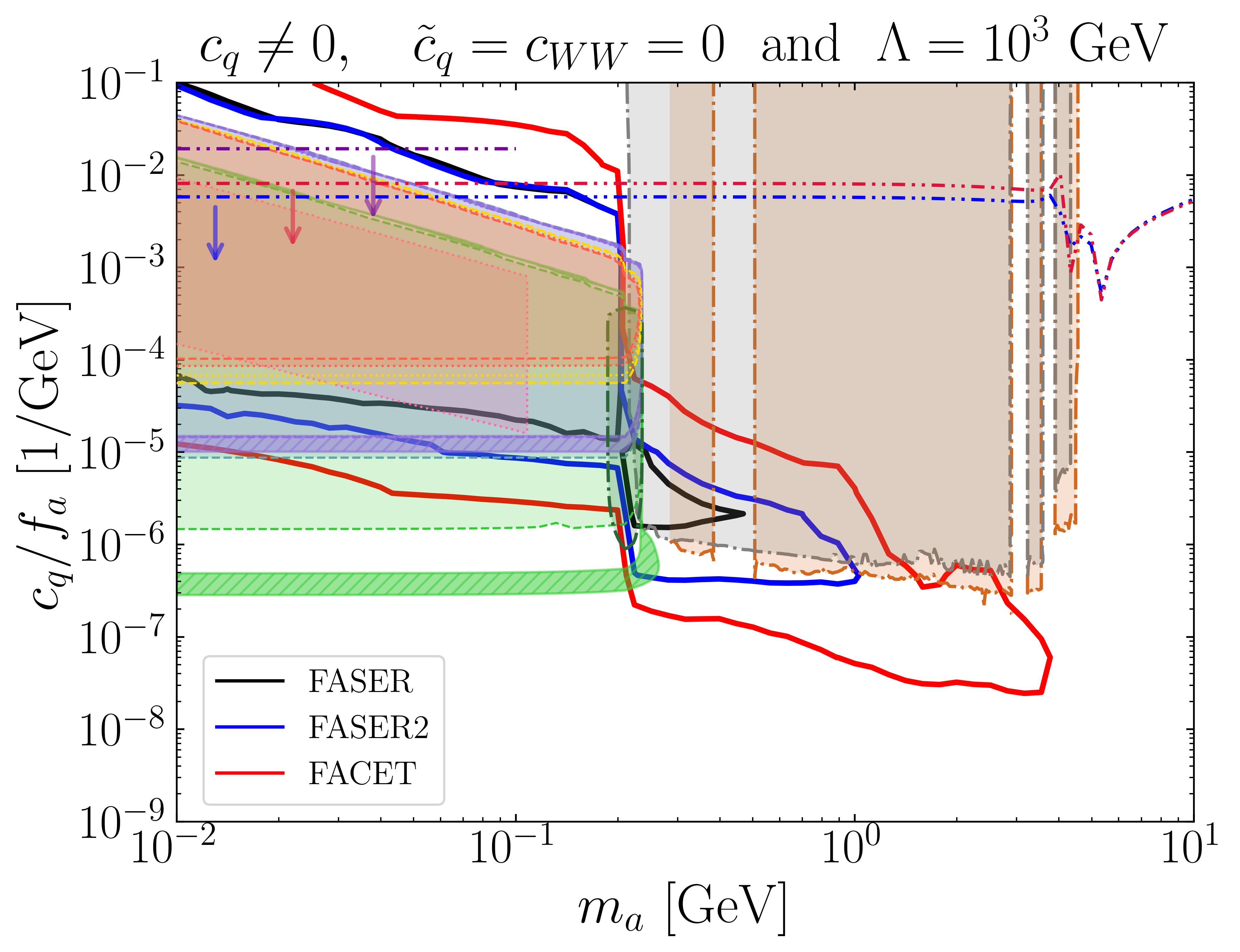}
\includegraphics[width=0.475\textwidth]{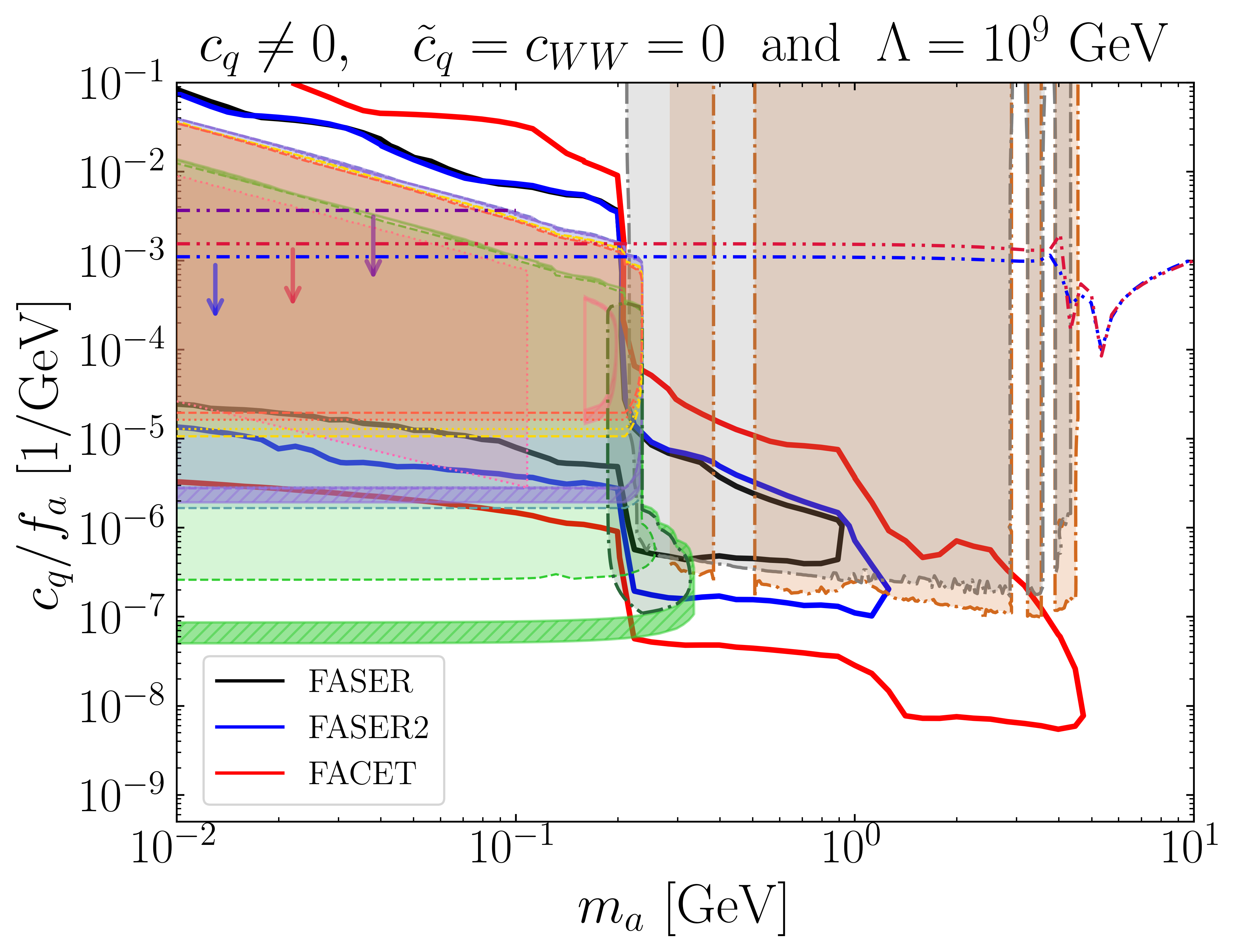}\\
\includegraphics[width=0.475\textwidth]{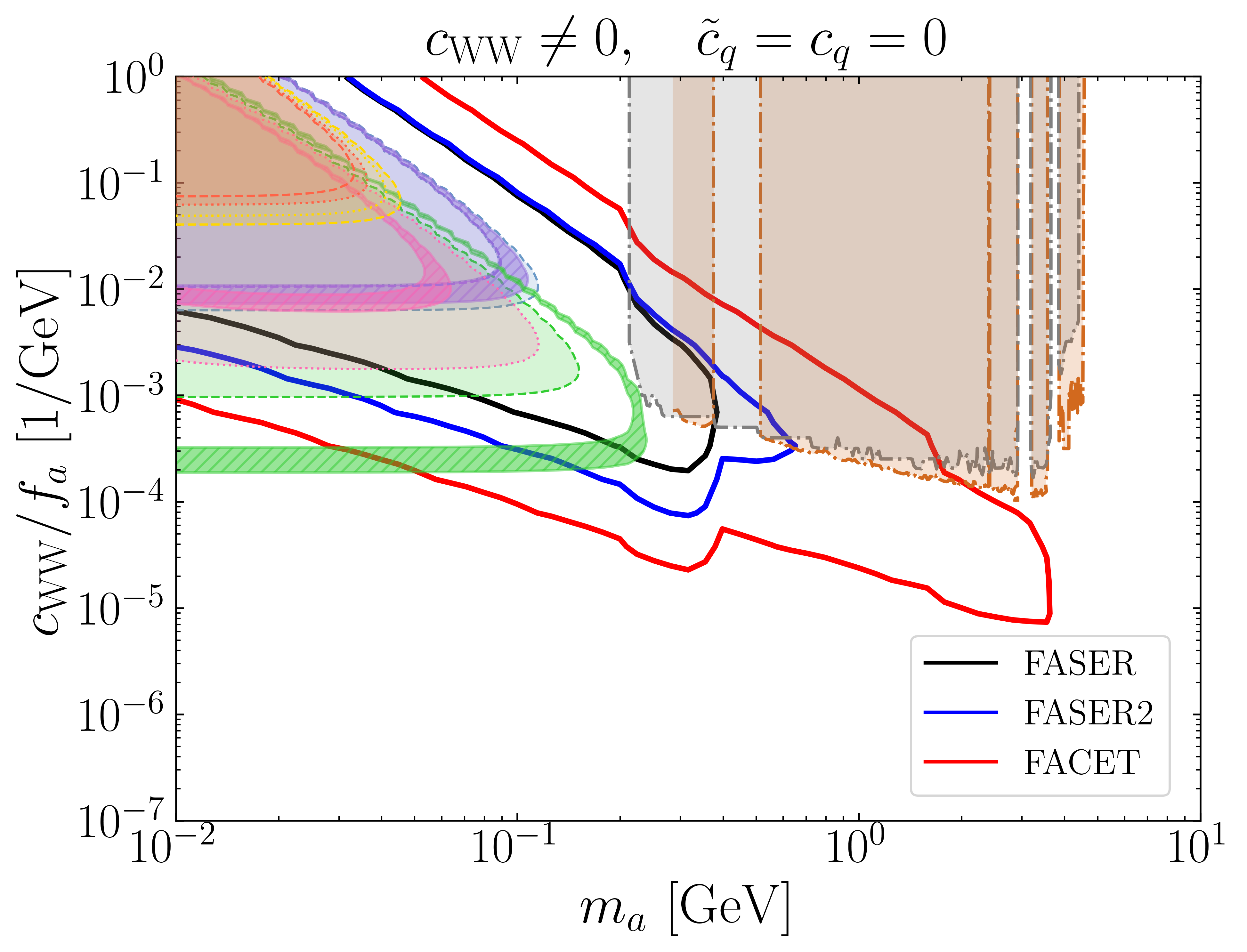}
\includegraphics[width=0.475\textwidth]{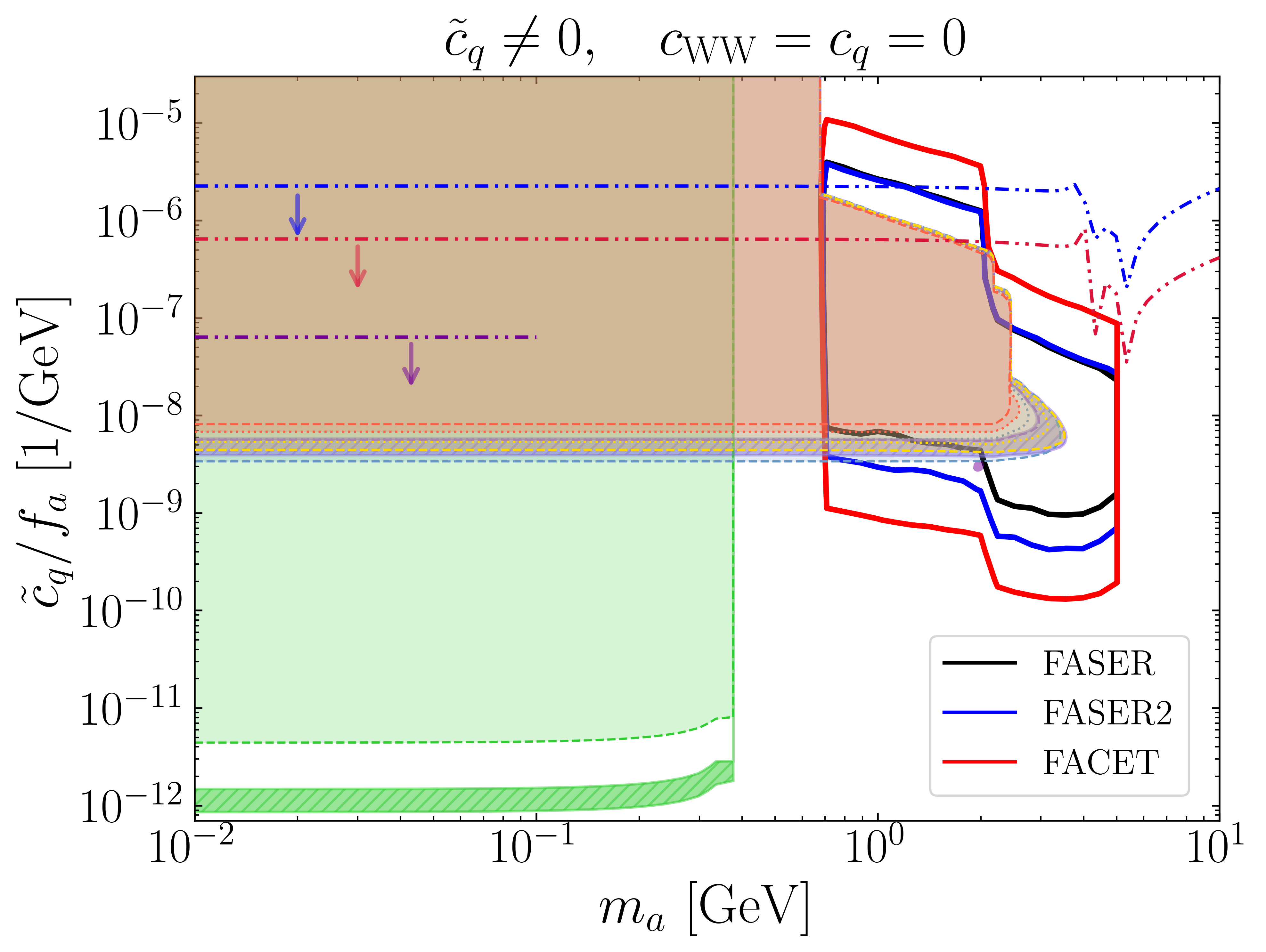}
\end{center}
\caption{The sensitivities of FASER (black), FASER2 (blue) and FACET (red) to ALP for the case $c_q\neq0,~c_{WW}=\tilde{c}_q=0$ with UV scale $\Lambda=10^3$ GeV (top left) and $\Lambda=10^9$ GeV (top right), case $c_{WW}\neq0,~c_q=\tilde{c}_q=0$ (bottom left), case $\tilde{c}_q\neq0,~c_q=c_{WW}=0$ (bottom right). The reachable contours are shown by solid lines.
The flavor constraints from $B$ or $K$ decays are displayed using exclusion regions enclosed by dashed (decay to neutral mesons) or dotted (decay to charged mesons) lines.
The preferred region by $B^+\to K^++{\rm inv}.$ ($K^+\to \pi^++{\rm inv}.$) [$K_L\to \pi^0 \gamma\gamma$] is shown by a purple (green) [magenta] meshed band. For $B^+\to K^++{\rm inv}.$ in particular, the light and dark purple regions represent the constraints from non-fitted~\cite{He:2022ljo} and fitted~\cite{Fridell:2023ssf} results, respectively. The regions pointed to by the arrows are within $1\sigma$ preferred parameter space by $K$ or $B$ meson mixing.
}
\label{fig:cma}
\end{figure}

In Fig.~\ref{fig:cma} we show the sensitivity of
FASER (black), FASER2 (blue) and FACET (red) to individual ALP couplings as a function of the ALP mass $m_a$. The projection contours in the plane of $c_i/f_a$ versus $m_a$ with $c_i=c_q$ (top panels), $c_i=c_{WW}$ (bottom left) and $c_i=\tilde{c}_q$ (bottom right) are drawn under the assumption of the presence of only one coupling at a time.
As illustration, we assume two UV scales for the flavor-conserving quark coupling, i.e., $\Lambda=10^3$ GeV in the top-left panel of Fig.~\ref{fig:cma} and $\Lambda=10^9$ GeV in the top-right panel. The closed contours indicate the sensitivity reach of FASER, FASER2 and FACET.
An ALP with a coupling above the upper bound would decay before getting into the LLP detector. A coupling smaller than the lower bound makes the production rate too low to generate enough signal events. FASER can probe the quark coupling $c_q/f_a$ down to $10^{-6}~{\rm GeV}^{-1}$.
FASER2 and FACET extend the sensitivity to $10^{-7}~{\rm GeV}^{-1}$ and $10^{-8}~{\rm GeV}^{-1}$, respectively.
For $m_a\lesssim m_K$ the dominant production mode is $K$ meson decay, while $B$ meson decays are most important for larger ALP masses up to $m_a\sim 2m_\tau$. The sensitivity region broadens for larger scales $\Lambda$, because the low-scale quark coupling $c_q(\mu<\mu_w)$ increases by a few times when running down from a higher UV scale. This is illustrated in the top two figures.
For the pure-gauge coupling case with non-zero $c_{WW}$, FASER (FASER2) [FACET] can reach a much higher limit of $c_{WW}/f_a\sim 10^{-3}~{\rm GeV}^{-1}$ ($10^{-4}~{\rm GeV}^{-1}$) [$10^{-5}~{\rm GeV}^{-1}$] because the ALP couplings are only induced at loop-level in this case. For the case of only tree-level couplings induced by $c_{\tilde{q}}$, the forward facility is only sensitive to ALP mass with $m_a>m_K + m_\pi\sim 0.6~{\rm GeV}$ from $B$ meson decay. The reachable bounds are much lower, extending to the level of $\tilde{c}_q/f_a\sim 10^{-9}-10^{-10}~{\rm GeV}^{-1}$, because the flavor-violating coupling is not loop-suppressed.

The flavor constraints are also displayed in Fig.~\ref{fig:cma}. The preferred region by $B^+\to K^++{\rm inv}.$ ($K^+\to \pi^++{\rm inv}.$) [$K_L\to \pi^0 \gamma\gamma$] is shown by a purple (green) [magenta] meshed band. For $\Lambda=10^3$ ($10^9$) GeV, the favored value of $c_q/f_a$ becomes $10^{-5}~{\rm GeV}^{-1}$ ($2\times 10^{-6}~{\rm GeV}^{-1}$) and $3\times 10^{-7}~{\rm GeV}^{-1}$ ($8\times 10^{-8}~{\rm GeV}^{-1}$) from $B^+\to K^++{\rm inv}.$ and $K^+\to \pi^++{\rm inv}.$, respectively.
For $B^+\to K^++{\rm inv}.$ in particular, the dark purple region represents the constraint from the fitted result~\cite{Fridell:2023ssf}. One can see that for $m_a\approx 2$ GeV, only $\tilde{c}_q/f_a\sim 3\times 10^{-9}~{\rm GeV}^{-1}$ is favored by a narrow region in the bottom-right plot. For the flavor-conserving couplings $c_q$ and $c_{WW}$, ALPs with a mass of $m_a\gtrsim 2m_\mu$ is short-lived and visibly decays inside the Belle II detector.
Other constraints are shown by enclosed exclusion regions.

The majority of the parameter space accessible to forward facilities has already been ruled out by constraints from heavy mesons' invisible decays for $m_a\lesssim 0.2~{\rm GeV}$. The kaon's invisible decay provides the most severe constraint. For heavier mass region, the invisible decay rate is sufficiently suppressed by the exponential factor ${\rm exp}(-L/L_a)$. The visible decays $B\to K\mu^+\mu^-$ exclude the region of $c_q/f_a\gtrsim 10^{-7}~{\rm GeV}^{-1}$ and $0.2~{\rm GeV}\lesssim m_a\lesssim m_B-m_K$. Only a limited region of the parameter space remains viable for exploration by FACET.
The regions pointed to by the arrows are within $1\sigma$ preferred parameter space by $K$ or $B$ meson mixing in Fig.~\ref{fig:cma}. Here we add the uncertainties of SM prediction and experimental measurement in quadrature. For $K$ meson mixing, we follow the approach using Chiral Perturbation Theory (ChPT) in Ref.~\cite{MartinCamalich:2020dfe}. As ChPT is applicable to very light ALP, we restrict the constrained parameter space to $m_a<100$ MeV. As the experimental measurements are in good agreement with SM predictions, arbitrarily small coupling values satisfy the $1\sigma$ constraints. For $c_{WW}$ inducing flavor-violating coupling at loop level, in particular, the whole parameter space is allowed by neutral meson mixing measurements and thus we do not show its preferred region.

\begin{figure}[tbp!]
\begin{center}
\includegraphics[width=0.475\textwidth]{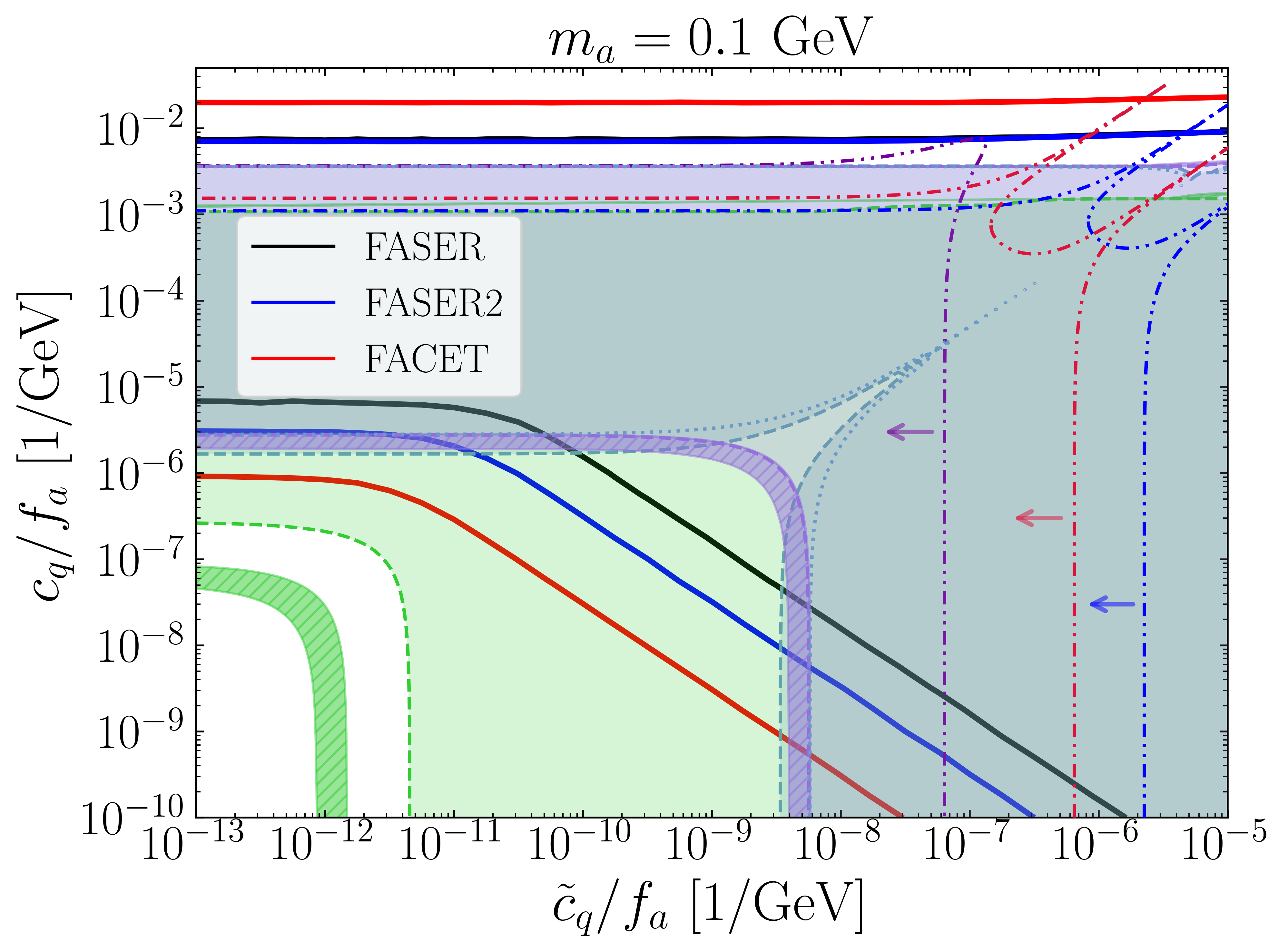}
\includegraphics[width=0.475\textwidth]{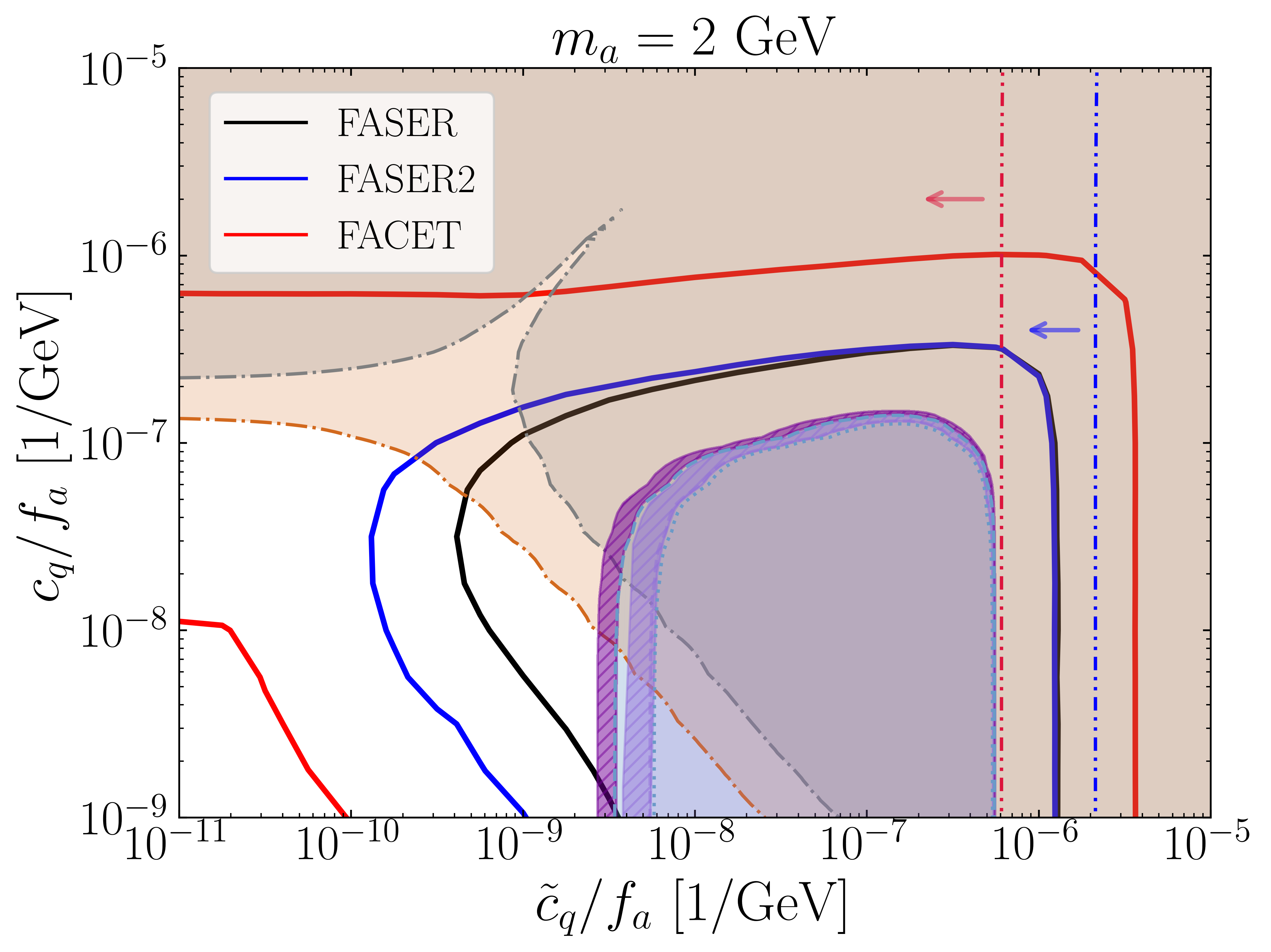}\\
\includegraphics[width=0.475\textwidth]{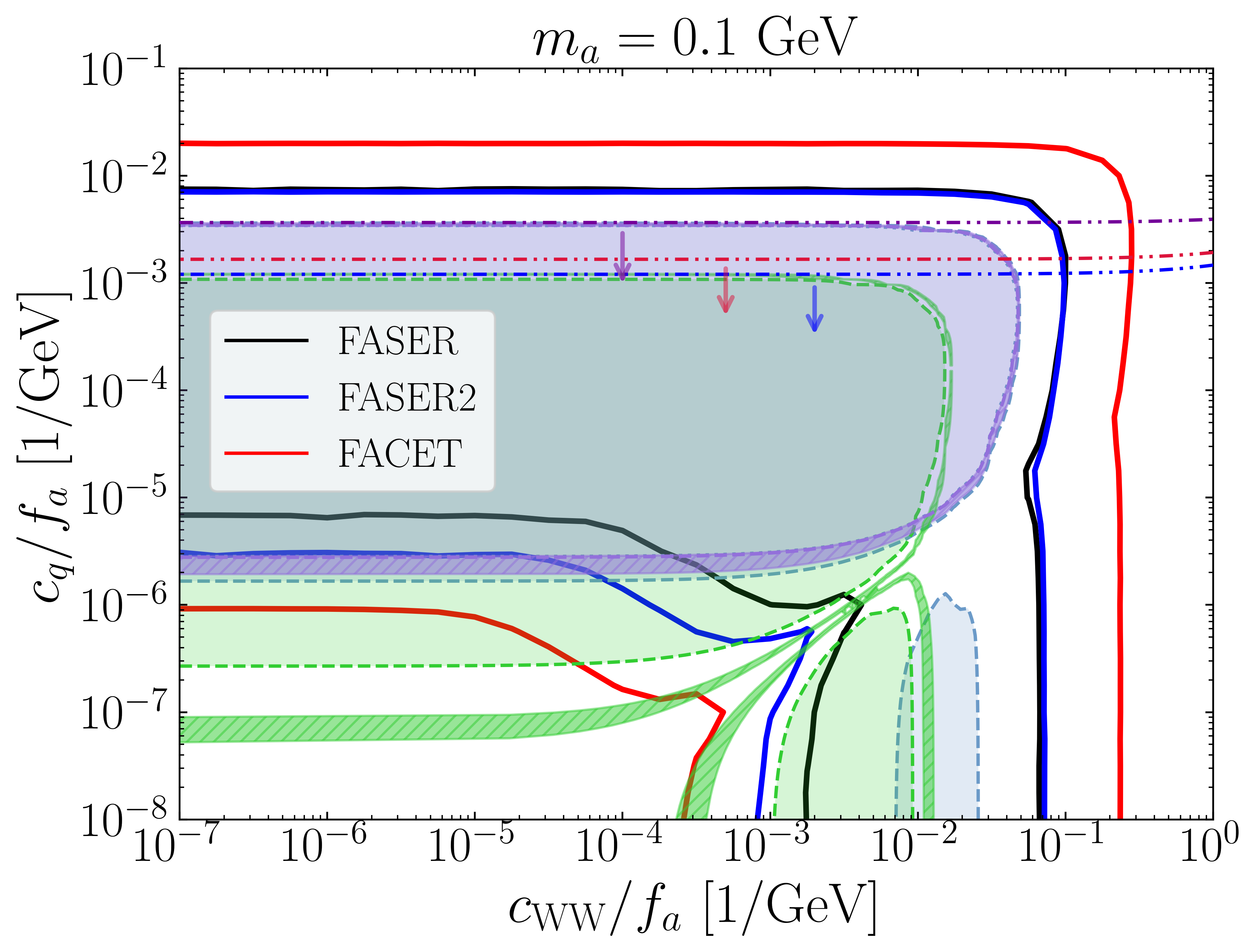}
\includegraphics[width=0.475\textwidth]{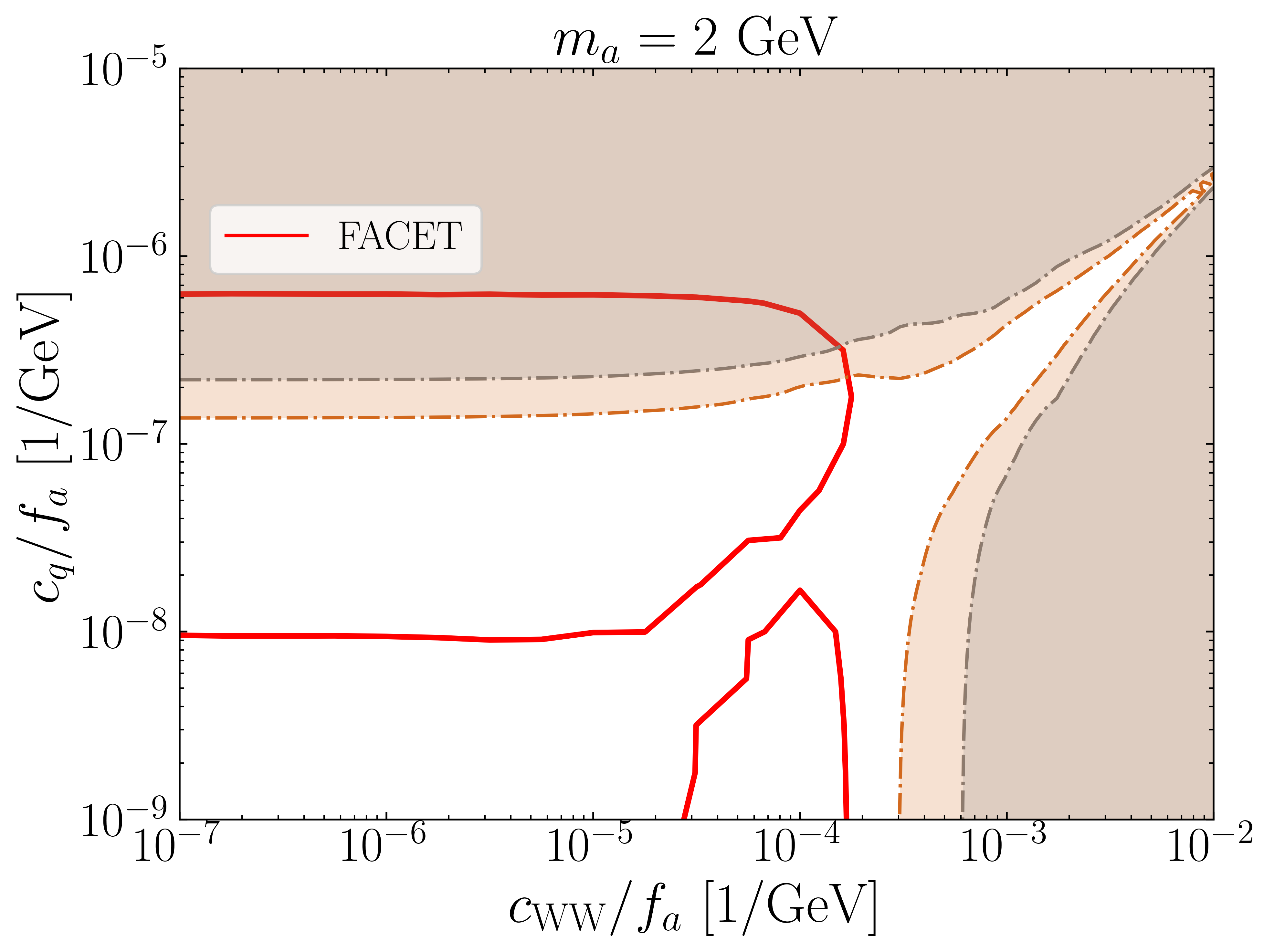}
\end{center}
\caption{The sensitivity of FASER (black), FASER2 (blue) and FACET (red) to the ALP couplings $c_q/f_a$ versus $c_{\tilde{q}}/f_a$ (top) and $c_q/f_a$ versus $c_{WW}/f_a$ (bottom) with UV scale $\Lambda=10^9$ GeV. For the ALP mass, two benchmarks $m_a=0.1$ GeV (left) and 2 GeV (right) are considered. The most stringent constraints from $B\to K^{(\ast)}+{\rm inv}.$ (purple and blue regions), $B\to K^{(\ast)}\mu^+\mu^-$ (gray and brown regions), $K_L\to \pi^0 +{\rm inv}.$ (green region), and the preferred regions from $B^+\to K^+ + {\rm inv}.$ (purple meshed band) and $K^+\to \pi^+ + {\rm inv}.$ (green meshed band) are also shown, as labeled in Fig.~\ref{fig:cma}. For $B^+\to K^++{\rm inv}.$ in particular, the light and dark purple regions represent the constraints from non-fitted~\cite{He:2022ljo} and fitted~\cite{Fridell:2023ssf} results, respectively. The $1\sigma$ preferred regions by $K$ or $B$ meson mixing are arrowed in the parameter space.
}
\label{fig:cqt-cff}
\end{figure}

Next we display the correlation of two ALP couplings in the sensitivity of accelerator experiments. The panels on the top of Fig.~\ref{fig:cqt-cff} show the reachable region in the plane of $c_q/f_a$ versus $c_{\tilde{q}}/f_a$ with UV scale $\Lambda=10^9$ GeV. Two benchmarks $m_a=0.1$ GeV (left panel) and 2 GeV (right panel) are considered for the ALP mass. Generally, there is an upper bound on each coupling because the ALP with the coupling beyond that maximum is not sufficiently long-lived and cannot reach the detector. The only exception is the flavor-violating coupling $c_{\tilde{q}}/f_a$ in the case of light ALP, e.g. $m_a=0.1$ GeV as shown in the top-left panel of Fig.~\ref{fig:cqt-cff}. In this case, the main ALP production process comes from $K$ meson decay and the light ALP can hardly decay into flavor-violating quark pairs due to kinematics. Thus, the ALP lifetime does not depend on $c_{\tilde{q}}$ and it would be considered as an LLP for arbitrarily large value of $c_{\tilde{q}}/f_a$ as long as $c_q/f_a$ is small enough. For the case of $m_a=2$ GeV in the right panel, the ALP can decay into flavor-violating quark pairs $sd$ and $cu$ through the $c_{\tilde{q}}$ coupling. There appears an upper reach limit on $c_{\tilde{q}}/f_a$ as well. Moreover, the tree-level coupling $c_{\tilde{q}}$ contributes to the production rate more significantly than the flavor-conserving coupling $c_q$ through a loop. Thus, the sensitivity of forward facilities depends on $c_{\tilde{q}}/f_a$ more significantly than $c_q/f_a$. We also show the most stringent constraints from $B\to K^{(\ast)}+{\rm inv}.$ (purple and blue regions), $B\to K^{(\ast)}\mu^+\mu^-$ (gray and brown regions), $K_L\to \pi^0 +{\rm inv}.$ (green region), and the preferred regions from $B^+\to K^+ + {\rm inv}.$ (purple meshed band) and $K^+\to \pi^+ + {\rm inv}.$ (green meshed band). The $1\sigma$ preferred regions by $K$ or $B$ meson mixing are arrowed in the parameter space. The forward facilities can probe viable regions of small couplings for relatively heavy ALP.

The correlation of $c_q/f_a$ versus $c_{WW}/f_a$ is displayed in the bottom panels of Fig.~\ref{fig:cqt-cff}, for $m_a=0.1$ GeV (left panel) and $m_a=2$ GeV (right panel).
Although for $m_a=0.1$ GeV case there is no ALP decay to flavor-violating quark pairs with $c_{WW}$ involved, the gauge coupling $c_{WW}$ contributes to $a\to \gamma\gamma, \ell^+\ell^-, q\bar{q}$ decays through one-loop diagrams. There is thus a weak upper bound on $c_{WW}/f_a$. While for the $m_a=2$ GeV case the reachable limits are stronger than those in the case of light ALP. Moreover, only FACET is sensitive to this large mass case as $c_{WW}/f_a$ does not contribute much. Note that the $c_q$ and $c_{WW}$ couplings are both involved in the $G_{d_i d_j}$ coupling in Eq.~(\ref{eq:Gij}). There may appear a large cancellation between the two couplings in the production rates and thus the ALP searches loose sensitivity along this direction
in the parameter space of these two couplings. This cancellation may also happen in the parameter space of $c_q/f_a$ and $c_{\tilde{q}}/f_a$ as shown in the top panels of Fig.~\ref{fig:cqt-cff}.

\section{Conclusion}
\label{sec:Con}

The search for ALP flavor-changing couplings requires rather different experimental strategies and facilities. In this work, we revisit flavor constraints on the feeble interactions of light ALP in light of the recent measurements of rare decays $B\to K + X$ and $K\to \pi +X$. With improved ALP decay calculation, we also investigate the sensitivity of forward accelerator experiments (FASER, FASER2 and FACET) to long-lived ALP with small mass.

We incorporate tree-level quark flavor-violating couplings to compute quark flavor-violating processes arising from both fermionic ALP couplings and electroweak ALP couplings, while accounting for the RG effects from high-energy scale to low-energy scale. The decay rates of ALP and heavy meson to light ALP are then numerically computed. We use the package FORESEE to evaluate the sensitivity reach of forward facilities to feeble couplings of a long-lived light ALP.
Our analysis of a quark flavor-violating light ALP goes beyond previous work by including the complete hadronic decay modes in the calculation of ALP decay rate and exclusive heavy meson decays as ALP production processes by following the procedure in Ref.~\cite{DallaValleGarcia:2023xhh}.

Our results demonstrate the complementarity of flavor experiments and the forward accelerator facilities for the probe of a light ALP. For $m_a \lesssim 0.2$ GeV most of the parameter space accessible to forward facilities has already been ruled out by constraints from heavy meson decays. For $m_a \gtrsim 0.2$ GeV forward accelerator experiments are able to probe smaller ALP couplings which are unexplored by flavor experiments. In absence of flavor-violating ALP couplings at the UV scale, FASER probes ALP-quark couplings down to $10^{-6} \,\mathrm{GeV}^{-1}$, while FASER2 and FACET extend the sensitivity by 1 and 2 orders of magnitude, respectively. For tree-level quark flavor-violating couplings at the UV scale, the forward accelerator experiments reach a sensitivity of $10^{-9} - 10^{-10}\, \mathrm{GeV}^{-1}$ and thus are able to probe even smaller couplings. The forward accelerator experiments are also able to probe the region of parameter space preferred by the recent measurement of $B^+\to K^+ +\mathrm{inv}.$~\cite{Belle-II:2023esi} and thus are able to test the ALP explanation of the observed excess.

\section*{ACKNOWLEDGMENTS}
We would like to thank Maksym Ovchynnikov for the reply regarding to SensCalc package.
T.~L. is supported by the National Natural Science Foundation of China (Grant No. 12375096, 12035008, 11975129).
M.~S. acknowledges support by the Australian Research Council Discovery Project DP200101470.

\bibliography{refs}

\end{document}